\documentclass[10pt,a4paper]{article}
\usepackage{a4wide}
\usepackage[boldonly]{chemie}
\makeatletter \renewcommand\@biblabel[1]{(#1)} \makeatother
\usepackage{amssymb}
\makeatletter
\usepackage{graphicx}
\usepackage{hyperref}
\newcommand{\mcc} [2]{\multicolumn{#1}{c}{#2}}
\newcommand{\mcl} [2]{\multicolumn{#1}{l}{#2}}
\newcommand{\mclb}[2]{\multicolumn{#1}{l}{\mbox{\sffamily\small\rule[-.5em]{0em}{2em}#2}}}
\newcommand{\Graphics}[2][scale=1]{%
 \def\blafile{ps/#2}
    \includegraphics[#1]{\blafile}}
\makeatother
\newcommand{\mb}{\mbox{-}}
\newsavebox{\BoxBox}
\sbox{\BoxBox}{\rule{1ex}{1ex}}

\newbox{\vers}
\title{Hydration of polyelectrolytes studied by molecular dynamics simulation}
\author{%
  Oliver Biermann$^\ast$  \and 
  \href{http://www.basf.de}{Erich H\"adicke$^\dag$}        \and
  \href{http://www.basf.de}{Sebastian Koltzenburg$^\dag$}  \and
  \href{http://www.basf.de}{Michael Seufert$^\dag$}        \and
  \href{http://www-theory.mpip-mainz.mpg.de/~mplathe}%
  {Florian M\"uller-Plathe$^\ast$} \and
  \begin{minipage}{.9\linewidth}
  \small
  $^\ast$%
  Max-Planck-Institut f\"ur Polymerforschung,
  Ackermannweg~10, 55128 Mainz, Germany\\
  $^\dag$ BASF AG, 67056 Ludwigshafen, Germany\\
  \small to be submitted to Macromolecules
 \end{minipage}
}

\begin{document}
\maketitle
 \begin{abstract}\small
   Molecular dynamics simulations of diluted ($\approx 2.5$ weight percent)
   aqueous solutions of two polyelectrolytes, namely sodium carboxy methyl
   cellulose (CMC) and sodium poly(acrylate) (PAA) have been performed.
   Water and counterions were taken into account explicitly.
   
   For CMC the substitution pattern and starting conformation is
   all-important.  Two simulations of CMC oligomers resulted in different
   structures: One molecule takes a stretched conformation, while the second
   one keeps a globule-like, toroidal one.  PAA is stretched during the whole
   simulation, with an average characteristic ratio of $8.3$.
   
   On a local atomistic scale CMC and PAA have different hydrogen-bond
   properties.  The COO$^-$ groups of PAA can only act as hydrogen bond
   acceptors, but due to the high negative charge density there are
   still more water molecules assembled around PAA than around CMC.  There
   are $0.036$\,bonds/amu respectively $0.029$\,bonds/amu to water for
   the two CMC oligomers, but more than twice as many for PAA:
   $0.083$\,bonds/amu.  Beside intermolecular hydrogen bonding, there is
   a significant amount of intramolecular H-bonding for CMC, which is
   influenced by the COO$^-$ groups, which act as strong H-acceptor.  In
   contrast to hydroxy- and carboxylic groups, ether oxygens are hardly
   involved into hydrogen bonding.
 \end{abstract}  

%%%%%%%%%%%%%%%%%%%%%%%%%%%%%%%%%%%%%%%%%%%%%%%%%%%%%%%%%%%%%%%%%%%%%%
%%
%%
%%
%%
%%
%%%%%%%%%%%%%%%%%%%%%%%%%%%%%%%%%%%%%%%%%%%%%%%%%%%%%%%%%%%%%%%%%%%%%%

 \section{Introduction}
  
 Polyelectrolytes play an important role in industrial chemistry.  The fields
 of application range from tailormade thickeners to paper finishing or ore
 preparation. Polysaccharide derivatives represent one interesting class of
 polyelectrolytes.  In particular, cellulose products are important compounds.
 For our simulation study carboxy methyl cellulose (CMC) is chosen as an
 example for a polyelectrolyte derived from a natural polymer.
 Aqueous CMC solutions exhibit valuable properties, like a wide range of
 viscosity, non-toxicity and biodegradability.  Particularly for the
 high-purity consumer-product market (cosmetics, food stuffs), CMC is used.
 However, pricing becomes more important in bulk applications (clay and ore
 treatment, oil-drilling). Hence it is desirable to replace some of the
 high-cost--high-selective chemicals with low-cost equivalents, like
 poly(acrylic acid) (PAA), which is the prototype of industrial synthetic
 polyelectrolytes.  PAA is the other polymer studied in this work.
 
 Most published work on aqueous CMC and PAA solutions was done experimentally
 using chromatography~\cite{heinze97a}, ${}^{13}$C~nuclear magnetic
 resonance~\cite{baar94,chaudhari87} and rheological
 techniques~\cite{kaestner97}.  Theoretical approaches are scarce.  We are
 aware of only one paper~\cite{davis91}, which treats CMC by the
 worm-like-chain theory.  This electrostatic theory successfully rationalizes
 some of the global properties of CMC, but as a rather generic approach it
 does not allow for detailed predictions on an atomistic time and length
 scale.  Similar restrictions apply also to Monte Carlo simulations of
 polyelectrolyte chains in a cell model, where the solvent is treated as a
 dielectric continuum~\cite{ullner98}.  Especially local interactions hydrogen
 bonds (hydrogen bridges) are neglected in theories and non-atomistic
 simulations.  With two or three hydrogen-bond donor groups per repeat unit
 and even more acceptors sites (including charged COO$^-$ groups), this type
 of interaction is likely to be very important for the behaviour of CMC in
 water.  Experimental techniques, on the other hand, suffer from different
 problems: NMR provides averaged local properties.  Rheology derives and
 verifies scaling laws, but different polyelectrolytes lose their chemical
 ``identity'' and show the generic behaviour of excluded-volume chains.
 Atomistic molecular dynamics (MD) simulations cannot overcome all these
 problems, but they can provide some more detailed information.
 
 The aim of this study is to understand better the structural and dynamic
 aspects of the hydration of CMC and PAA and to compare the two polymers.  To
 this end, we investiagate both the chain properties and the interaction of
 chains with their immediate solvent environment.  Atomistic simulation is
 confined to the study of small system sizes.  However, in combination with
 coarse-graining methods~\cite{baschnagel00,tschoep98,tschoep98b} even some
 mesoscopic properties may be explored.  Thus, a second goal of this study was
 to produce atomistic structural information, from which
 coarse grained models of e.\,g. PAA can be generated.
 
%%%%%%%%%%%%%%%%%%%%%%%%%%%%%%%%%%%%%%%%%%%%%%%%%%%%%%%%%%%%%%%%%%%%%%
%%
%%
%%
%%
%%
%%%%%%%%%%%%%%%%%%%%%%%%%%%%%%%%%%%%%%%%%%%%%%%%%%%%%%%%%%%%%%%%%%%%%%

 \section{Computational Details}\label{sec:CompDeta}
 
\begin{figure}
   \[   \Graphics[scale=.6]{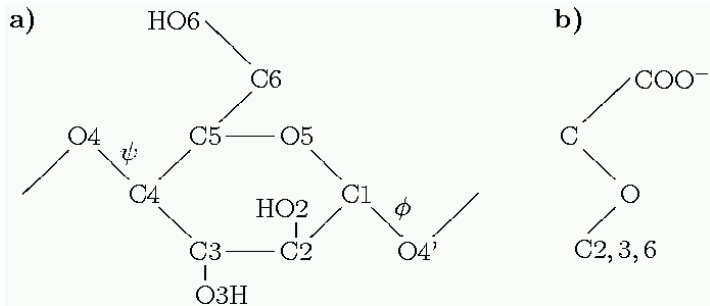}  \]
   \caption{%
       \emph{a)} Nomenclature of an anhydroglucose unit (AGU).
       \emph{b)}A carboxy methyl group, which may be attached to C2, C3 or C6. 
          Aliphatic hydrogens are omitted for clarity.}
   \label{fig:AGUI}
 \end{figure}

 The exact molecular composition of CMC is a consequence of the molecule's
 production history. Usually OH-groups are randomly substituted by carboxy
 methyl units~\cite{baar94,heinze93} and the degree of substitution (DS) for
 most commercial CMC's lies in the range DS=$0.6$--$1.2$ substitutions per
 anhydroglucose unit (AGU).  There is evidence for a preferential substitution
 during heterogeneous, wet conversion of cellulose to CMC: Heinze and
 coworkers used liquid chromatography to analyse the substitution pattern of
 Sodium-CMC. Their analysis was preceded by hydrolysis of the glucosidic link,
 by which information on the ring-substitution sequence is lost.  For the AGU,
 the following substitution statistics was found~\cite{heinze96c}: 23\Percent\ 
 of all O2-oxygens were substituted, followed by the O6 (15\Percent) and O3
 with 11\Percent\ of all respective atoms.  The glucose nomenclature is shown
 in figure~\ref{fig:AGUI}.  The diversity of molecular substitution patterns
 of CMC makes it difficult to model such a compound by an oligomer, because
 there is no ``typical'' molecule or structure.
 
 To have at least some variety, two CMC oligomers were simulated.
 Substitution patterns were generated at random, with weights for the three
 exocyclic oxygens ($P(O2)=0.23$, $P(O3)=0.12$, $P(O6)=0.15$).  They are
 summarized in table~\ref{tab:cmcsubst}.  The first molecule (CMC~I) consists
 of seven AGUs and is substituted five times, the second one (CMC~II) consist
 of eight AGUs with six OH groups substituted.  The degrees of substitution
 are DS=$0.70$ and DS=$0.75$, respectively.  Due to three sites, at which
 substitution can take place (neglecting the chain ends) there are eight
 different possible substitution patterns on a single AGU.  Four of these
 (O2\,O3, O3, O2, O6) are present in the first simulated CMC oligomer and one
 further pattern (O3\,O6) in the second CMC oligomer.

 \begin{table}
   \[
   \begin{tabular}{*{9}{l}}
        & \mcc8{Anhydroglucose unit number}               \\\cline{2-9}
     \rule{0pt}{1em}%
        &  1  &  2  &  3  &  4  &  5  &  6  &  7   & (8)  \\
     O2 &     &     &  I  &     &  II &I, II&      &      \\
     O3 &     &  II &  I  &     &  I  &     & II   &      \\
     O6 &     &     &     &     &     &     &I, II &  II  \\
   \end{tabular}
   \]
   \caption{Substitution pattern of both simulated carboxy methyl cellulose
     (CMC)  molecules.
     An entry in the table for one of the CMC molecules (I or II) indicates a
     substitution at the respective OH-Group (O2, O3 and O6).  Anhydroglucose units
     are numbered from 1 to 7 (CMC~I) or 1 to 8 (CMC~II).}%
   \label{tab:cmcsubst}
 \end{table}
 
 All carboxylic acid groups were assumed to be dissociated, neutrality
 of the systems was maintained by five, respectively six sodium
 counterions.  As the carboxylic acid group has a $\mathrm{pK_A}$ of
 about $4$, the system is basic.  Molecular weights are $1440$\,amu
 (seven~AGUs, CMC~I) and $1660$\,amu (eight~AGUs, CMC~II), respectively.
 Molecule CMC~I was solvated in 3789, molecule~II in 3917 water
 molecules.  The concentrations are about $2.1$ and $2.4$\,weight
 percent (excluding the sodium-counterions).  Therefore both simulations
 were performed in the dilute regime (no added salt), where
 polyelectrolytes are expected to be stretched~\cite{kaestner97}.
 
 Polyacrylic acid (PAA) was modeled as one atactic oligomer strand, with $23$
 repeat units ($1636$\,amu) solvated in $3684$ water molecules (concentration
 excluding counterions $2.4$\, weight percent).  All COO$^-$ groups are negatively
 charged and there are $23$ sodium counterions.  Runs were performed under the
 same conditions as for CMC, including the GROMOS force
 field~\cite{vangunsteren:96} as detailed in table~\ref{tab:gromosNB}
 and~\ref{tab:gromosBND}.
 
 \begin{table}      % CMC ff part I
   \footnotesize
\renewcommand{\b}[1]{\underline{#1}}
$\begin{array}{*{6}{l}}
\mclb6{Nonbonded Parameters CMC} \\
\mbox{Name}&  \mbox{Type}        &  \sigma\unit*{nm} 
                                             & \varepsilon\unit*{kJ\,mol^{-1}}
                                                          &  q\unit*{e} 
                                                                     & \mbox m\unit*{amu}  \\
\mbox{C}   
 & \mathrm{COO^-}                &  0.33611  &  0.405870  &  0.2700  &  12.0100 \\ %&   11/1 \\ %   ce:
\mathrm{CH_1} 
 & \mathrm{O\mb \b{C1}\mb O}     &  0.38004  &  0.313940  &  0.4000  &  13.0190 \\ %&   12/1 \\ % ch1a:
 & \mathrm{\b{C2}\mb OH, \b{C3}\mb OH} 
                                 &  0.38004  &  0.313940  &  0.1500  &  13.0190 \\ %&   12/1 \\ % ch1b:
 & \mathrm{\b{C4}\mb OR_2, \b{C5}\mb C\mb OH}
                                 &  0.38004  &  0.313940  &  0.1600  &  13.0190 \\ %&   12/1 \\ % ch1c:
\mathrm{CH_2}  
 & \mathrm{R\mb CH_2\mb OH}\quad\mbox{and}  &&            &          &          \\ %&   13/1 \\ % ch2a:   
 & \mathrm{R\mb\b{CH_2}\mb O\mb CH_2\mb COO^-} 
                                 &  0.39199  &  0.489590  &  0.1500  &  14.0270 \\ %&   13/1 \\ % ch2b:
 & \mathrm{R\mb CH_2 \mb O\b{CH_2}\mb COO^-}\quad\mbox{and}
                                 &           &            &          &          \\ %&   13/1 \\ % ch2c:
 & \mathrm{R\mb O\mb \b{CH_2}\mb COO}^-
                                 &  0.39199  &  0.489590  &  0.2080  &  14.0270 \\ %&   13/1 \\ % ch2d:
\mbox{O}  
 & \mathrm{\mb O\mb H}           &  0.28706  &  1.010650  & -0.5480  &  15.9994 \\ %&    3/1 \\ %  oa:
 & \mathrm{\mb O\mb H~(chain~end)}
                                 &  0.28706  &  1.010650  & -0.5730  &  15.9994 \\ %&    3/1 \\ % oaa:
 & \mathrm{R\mb O\mb CH_2}\quad\mbox{and}  & &            &          &          \\ %&    3/1 \\ %  ob:
 & \mathrm{R\mb CH_2\mb \b{O}\mb CH_2\mb COO^-}
                                 &  0.28706  &  1.010650  & -0.3580  &  15.9994 \\ %&    3/1 \\ %  oc:
 & \mathrm{R\mb O\mb R~(ether)}  &  0.28706  &  1.010650  & -0.3600  &  15.9994 \\ %&    3/1 \\ %  od:  
\mbox{}
 &  \mathrm{\mb COO}^-           &  0.26259  &  1.725440  & -0.6350  &  15.9994 \\ %&    2/1 \\ % oe:
\mbox{H}
 &  \mathrm{\mb O\mb H}          &  0.00000  &  0.000000  &  0.3980  &   1.0080 \\ %&   18/1 \\ %    h:
\mbox{H}
 &  \mathrm{\mb O\mb H\;(chain~end)}
                                 &  0.00000  &  0.000000  &  0.3730  &   1.0080 \\ %&   18/1 \\ %   hb:
\mbox{OW}
 &  \mathrm{water}               &  0.31650  &  0.650300  &  -0.8200  &  15.9994 \\ %&    4/1 \\ %   ow:
\mbox{HW}
 &  \mathrm{water}               &  0.00000  &  0.000000  &  0.4100  &   1.0080 \\ %&   18   \\ %   hw:
\mbox{Na}^+
 &                               &  0.27300  &  0.358000  &  1.0000  &  22.9898 \\ %&   34/1 \\ %   na:
\mclb6{Additional Nonbonded Parameters PAA } \\
\mbox C & \mathrm{CH_1}          & 0.38004   & 0.31394    & 0.000    &  13.0190 \\
        & \mathrm{CH_2}          & 0.39199   & 0.48959    & 0.000    &  14.0270 \\
        & \mathrm{CH_3}          & 0.38750   & 0.73227    & 0.000    &  15.0350 \\
\end{array}$

%%% Local Variables: 
%%% mode: latex
%%% TeX-master: "paper"
%%% End: 

   \caption{Overview of all nonbonded force field parameter
            of our simulations.  A united-atom model is used, so no 
            explicit aliphatic hydrogens are present.  Lennard Jones and
            electrostatic interactions between atoms less than 3
            bonds apart are switched off. Lennard-Jones interactions between unlike
            atoms are treated using the Lorentz-Berthelot mixing
            rules~\cite{allen:96}. R refers to an aliphatic site.  Underlining
            is used to make the assignment unambiguous.}
            \label{tab:gromosNB}
 \end{table}

 \begin{table}      % CMC ff part II
   \footnotesize
   \begin{minipage}[t]{.49\linewidth}
$\begin{array}[t]{ll}
 \mclb2{Bond constraints CMC and PAS}     \\
    \mbox{Name}            & r_0\unit*{nm} 
                                       \\ %& \mbox{GROMOS bond type} \\
  \mathrm{CH_n\mb (COO)^-}     &  0.1530 \\ %& 26   \\
  \mathrm{CH_n\mb CH_{1|2}}  &  0.1520 \\ %& 25   \\
  \mathrm{CH_n\mb O}         &  0.1435 \\ %& 19   \\
  \mathrm{C\mb (OO)}         &  0.1250 \\ %&  5   \\
  \mathrm{O\mb H}            &  0.1000 \\ %&  1   \\
\end{array}$ 
%%% Local Variables: 
%%% mode: latex
%%% TeX-master: "paper"
%%% End: 
     \par
 $\begin{array}[t]{lll}
 \mclb3{Bond angles CMC and PAS} \\
 \mbox{Name}                       & \phi_0  & K_\phi    \\ %& \mbox{GROMOS}\\
                                   & \unit*{degree}& \unit*{kJ\,mol^{-1}\,rad^{-2}} \\ 
 \mathrm{CH_1\mb O\mb H}           &  109.5  &     450.0 \\ %& 11          \\
 \mathrm{CH_n\mb CH_n\mb CH_n}     &  109.5  &     285.0 \\ %& 7          \\
 \mathrm{CH_n\mb CH_n\mb O}        &  109.5  &     320.0 \\ %& 8          \\
 \mathrm{CH_n\mb O\mb CH_n}        &  109.5  &     380.0 \\ %& 9          \\
 \mathrm{O\mb CH_n\mb O}           &  109.5  &     285.0 \\ %& 8          \\
 \mathrm{CH_n\mb (COO)\mb (OO)}    &  117.0  &     635.0 \\ %& 21          \\
 \mathrm{(COO)\mb CH_n\mb O}       &  109.5  &     520.0 \\ %& 12          \\
 \mathrm{(OO)\mb (COO)\mb (OO)}    &  126.0  &     770.0 \\ %& 37          \\
 \mathrm{CH_2\mb CH_1\mb CH_2}     & 111.0   &     530.0 \\ % 14
\end{array}$
%%% Local Variables: 
%%% mode: latex
%%% TeX-master: "paper"
%%% End: 
     \par
$\begin{array}[t]{lll}
     \mclb3{Harmonic dihedrals CMC and PAS} \\
     \mbox{Name}                         &  \delta_0\unit*{degree}&   K_\delta\unit*{kJmol^{-1}rad^{-2}}\\
     \mathrm{C1\mb O5\mb O4\mb O2}   &  35.3 &   334.9 \\  % 2     1       6       114     2    
     \mathrm{C5\mb O5\mb C6\mb C4}   &  35.3 &   334.9 \\  % 2     5       6       7       4    
     \mathrm{C4\mb C3\mb O4\mb C5}   &  35.3 &   334.9 \\  % 2     4       3       8       5    
     \mathrm{C3\mb O3\mb C2\mb C4}   &  35.3 &   334.9 \\  % 2     3       10      2       4    
     \mathrm{C2\mb O2\mb C3\mb C1}   &  35.3 &   334.9 \\  % 2     2       9       3       1    
     \multicolumn{3}{l}{\mathrm{CH_2\mb (OO)\mb (OO)\mb (COO)}}  \\
                                        &  0.0      &   167.5 \\  %  40 42 43 41      
 \end{array}$
%%% Local Variables: 
%%% mode: latex
%%% TeX-master: "paper"
%%% End: 
   \end{minipage}
   \begin{minipage}[t]{.49\linewidth}
     \par
 $\begin{array}[t]{*{4}{l}}
  \mclb4{Torsional angles} \\
  \mbox{Name}                           & \tau_0 &  n & K_\tau  \\ %    & \mbox{GROMOS}\\ 
  \mclb4{CMC}                                                   \\
                                        &\unit*{degree} && \unit*{kJ\,mol^{-1}} \\
 \mathrm{H4\mb O4\mb C1\mb C2}      &     60.0 &  3 &  2.520  \\ %    &  12  \\
 \mathrm{O4\mb C1\mb C2\mb C3}      &     60.0 &  3 & 11.720  \\ %    &  17  \\
 \mathrm{O4\mb C1\mb C2\mb C3}      &     90.0 &  2 &  0.836  \\ %    &  7   \\
 \mathrm{O5\mb C1\mb C2\mb C3}      &     90.0 &  2 &  0.836  \\ %    &  7   \\
 \mathrm{O5\mb C1\mb C2\mb O2}      &     90.0 &  2 &  4.180  \\ %    &  8   \\
 \mathrm{O4\mb C1\mb C2\mb O2}      &     90.0 &  2 &  4.180  \\ %    &  8   \\
 \mathrm{C1\mb C2\mb O2\mb H2}      &     60.0 &  3 &  2.520  \\ %    &  12  \\
 \mathrm{C1\mb C2\mb C3\mb C4}      &     60.0 &  3 & 11.720  \\ %    &  17  \\
 \mathrm{C1\mb C2\mb C3\mb O3}      &     90.0 &  2 &  0.836  \\ %    &  7   \\
 \mathrm{O2\mb C2\mb C3\mb C4}      &     90.0 &  2 &  0.836  \\ %    &  7   \\
 \mathrm{O2\mb C3\mb C3\mb O3}      &     90.0 &  2 &  4.180  \\ %    &  8   \\
 \mathrm{C2\mb C3\mb O3\mb H3}      &     60.0 &  3 &  2.520  \\ %    &  12  \\
 \mathrm{C2\mb C3\mb C4\mb C5}      &     60.0 &  3 & 11.720  \\ %    &  17  \\
 \mathrm{C2\mb C3\mb C4\mb O4}      &     90.0 &  2 &  0.836  \\ %    &  7   \\
 \mathrm{O3\mb C3\mb C4\mb C5}      &     90.0 &  2 &  0.836  \\ %    &  7   \\
 \mathrm{O3\mb C3\mb C4\mb O4}      &     90.0 &  2 &  4.180  \\ %    &  8   \\
 \mathrm{C2\mb C1\mb O5\mb C5}      &     60.0 &  3 &  7.540  \\ %    &  14  \\
 \mathrm{C1\mb O5\mb C5\mb C4}      &     60.0 &  3 &  7.540  \\ %    &  14  \\
 \mathrm{C4\mb C5\mb C6\mb O6}      &     60.0 &  3 & 11.720  \\ %    &  17  \\
 \mathrm{C4\mb C3\mb C6\mb O6}      &     90.0 &  2 &  0.836  \\ %    &  7   \\
 \mathrm{O5\mb C5\mb C6\mb O6}      &     90.0 &  2 &  4.180  \\ %    &  8   \\
 \mathrm{C5\mb C6\mb O6\mb H6}      &     60.0 &  3 &  2.520  \\ %    &  12  \\
 \mathrm{C6\mb C5\mb C4\mb C3}      &     60.0 &  3 & 11.720  \\ %    &  17  \\
 \mathrm{O5\mb C5\mb C4\mb C3}      &     90.0 &  2 &  0.836  \\ %    &  7   \\
 \mathrm{C6\mb C5\mb C4\mb O4}      &     90.0 &  2 &  0.836  \\ %    &  7   \\
 \mathrm{O5\mb C5\mb C4\mb O4}      &     90.0 &  2 &  4.180  \\ %    &  8   \\
 \mathrm{C3\mb C4\mb O4\mb C1}      &     60.0 &  3 &  7.540  \\ %    &  14  \\
%%%%%%%%%%%%%%%%%%%%%%%%%%%%%%%%%%%%%%%%%%%%%%%%%%%%%%%%%%%%%%%%%%%%%%%             COO group
 \mathrm{C2\mb C3\mb O3\mb CH_2}     &     60.0 &  3 &  7.540  \\ % &  14  \\
 \mathrm{C3\mb O3\mb CH_2\mb (COO)}  &     60.0 &  3 &  7.540     \\ % &  14  \\
 \mathrm{O2\mb CH_2\mb (COO)\mb (OO)}&   0.0 &  6 &  2.000     \\ % &  20  \\
 \mathrm{O3\mb CH_2\mb (COO)\mb (OO)}&   0.0 &  6 &  2.000     \\ % &  20  \\
%%%%%%%%%%%%%%%%%%%%%%%%%%%%%%%%%%%%%%%%%%%%%%%%%%%%%%%%%%%%%%%%%%%%%%%             PAS
  \mclb4{PAA}                                                   \\
\mathrm{CH_3\mb CH_1\mb CH_2\mb CH_1}   & 60.0  & 3  & 11.720     \\ 
\mathrm{CH_1\mb CH_2\mb CH_1\mb CH2}    & 60.0  & 3  & 11.720     \\
\mathrm{CH_2\mb CH_1\mb CH_2\mb CH1}    & 60.0  & 3  & 11.720     \\
\mathrm{CH_{2|3}\mb CH_1\mb (COO)\mb (OO)}  & 30.0  & 6  &  2.000     \\
\end{array}$

%%% Local Variables: 
%%% mode: latex
%%% TeX-master: "paper"
%%% End: 

   \end{minipage}
%SUBMIT\setcounter{page}{50}
   \caption{Overview of bonding parameters of our simulations.  
            ``(COO)'' is carboxylic carbon, ``(OO)'' carboxylic oxygen.
            All other atom type names are self-explanatory (see
            figure~\ref{fig:AGUI}).  Analytical forms of force field term are
            as explained in ref.~\cite{mueller-plathe93}.}
   \label{tab:gromosBND}
 \end{table}
 
 To ensure a consistent forcefield, we used simple point charge (SPC)
 water~\cite{berendsen:81} with the GROMOS parameter set.  This combination
 has proven useful in several simulations of sugars in
 water~\cite{heiner98,mark94}.  Lennard-Jones interactions were truncated and
 shifted at a cutoff of $r_c=0.9\nm$. An isotropic pressure correction term
 was applied for site-site distances greater than the cutoff $r_c$.  Electrostatics
 were dealt with using the Coulomb-Potential with a reaction-field
 correction~\cite{neumann83,neumann80}:
 \[
  V_q(r_{ij}) = 
  \frac{q_i q_j}%
       {4\pi\epsilon\epsilon_0}
  \left(
    \frac{1}{r_{ij}}  +
    \frac{\epsilon_\mathrm{RF} - 1 }{2\epsilon_\mathrm{RF} + 1}
    \frac{r_{ij}^2}{r_c^3}
   \right)
 \]
 with the reaction field dielectric $\epsilon_\mathrm{RF}$ equal to the value
 for water ($\epsilon_\mathrm{RF}=78.5$).  All atoms -- except aliphatic
 hydrogens -- are explicitly modeled. In the GROMOS forcefield, aliphatic
 hydrogens are accounted for by a change of the parameters for the parent
 carbons.
 
 Bond lengths were held constant using the Shake
 procedure~\cite{berendsen:77}. Equations of motion were integrated using the
 leap-frog algorithm with a timestep of $2$\,fs. Other simulation parameters
 were a neighbor list, which was updated every $15$ steps with a list cutoff
 of $1\nm$, weak coupling to a temperature ($T=333.15\Kelvin$) and pressure
 ($p=1\unit{atm}$) bath~\cite{berendsen:84} with coupling times of $\tau_T =
 0.5\ps$ and $\tau_p = 2.5\ps$ (water compressibility
 $4.5\cdot10^{-10}\unit{kPa^{-1}}$).  For later analysis, coordinates were
 written every 195 steps. The overall simulation time was 4.5\,ns for CMC~I
 and $2.5$\,ns for CMC~II, each after $1$\,ns of equilibration.
 
 We used different starting geometries for both CMC~I and CMC~II.  The
 first one was started in a stretched conformation, the second, CMC~II,
 was prepared in a folded starting conformation: During the first $150\ps$ of
 equilibration, a globule-like conformation was enforced by five
 artificial harmonic bonds (force constant of
 $100\,\mathrm{kJ\,mol^{-1}nm^{-2}}$ with a minimum energy at $1\nm$
 between C1 atoms of different glucose rings).  PAA was started in the
 stretched (all-trans) conformation, equilibrated for
 $1$\,ns and simulated for about $4.5$\,ns.
 
 For analysis purpose, the center of mass of a repeat unit is caluclated
 using all its atoms, including any side groups if present.  The CMC
 repeat unit begins at C1 and ends at O4 according to the numbering in
 figure~\ref{fig:AGUI}.  The acrylate repeat unit coincides with the
 propylate unit $\mathrm{\mb{}CH_2\mb{}CH\mb{}COO^-}$.
 
 The simulations were done with our molecular dynamics package
 YASP, which is described in ref.~\cite{mueller-plathe93}.

%%%%%%%%%%%%%%%%%%%%%%%%%%%%%%%%%%%%%%%%%%%%%%%%%%%%%%%%%%%%%%%%%%%%%%
%%
%%
%%
%%
%%
%%%%%%%%%%%%%%%%%%%%%%%%%%%%%%%%%%%%%%%%%%%%%%%%%%%%%%%%%%%%%%%%%%%%%%

 \section{Results}

 \subsection{Global Chain Properties}
 
\begin{figure}[htbp]
   \[
   \Graphics[scale=.6,angle=0]{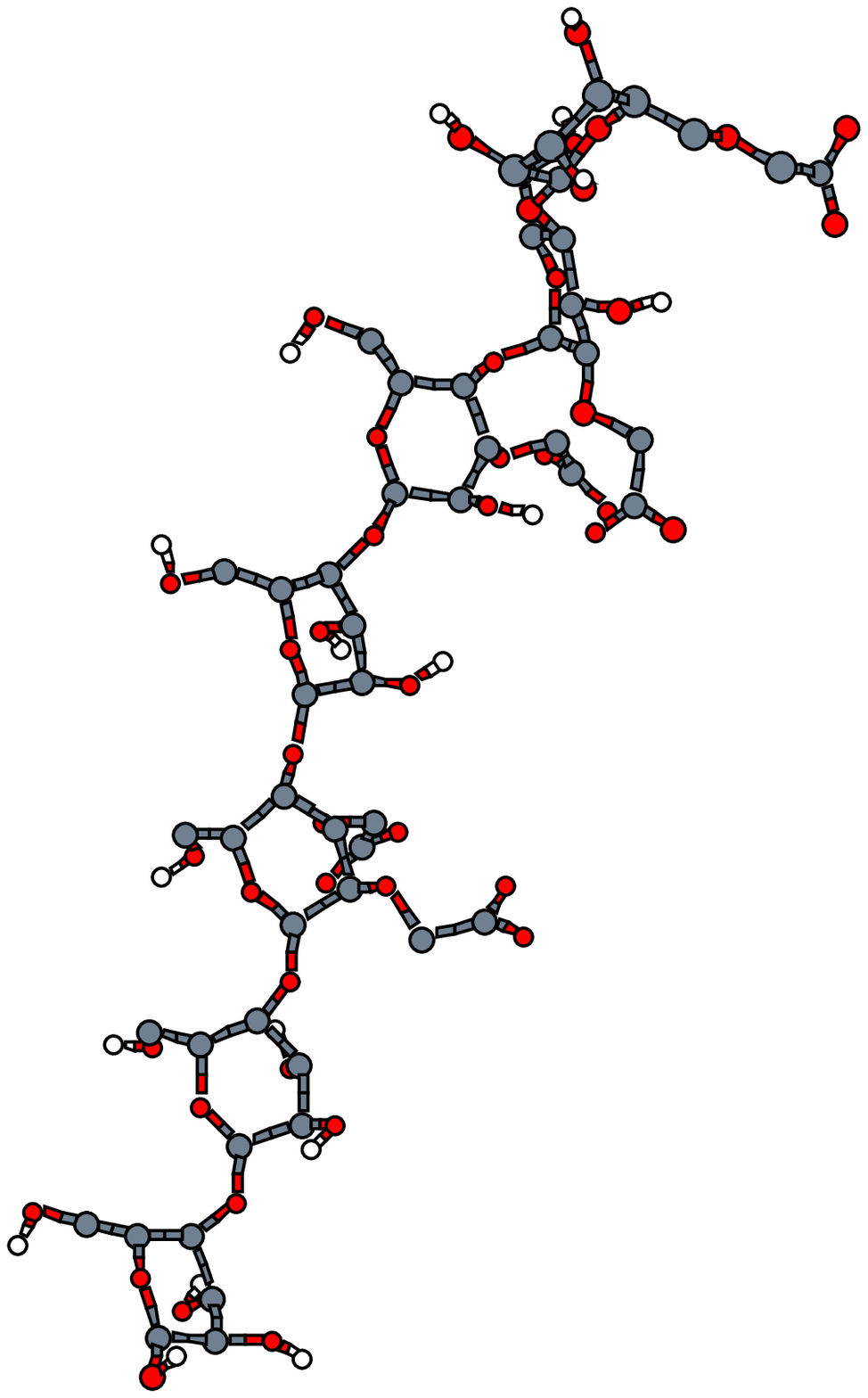}\hspace{-3cm}
   \Graphics[scale=.5]{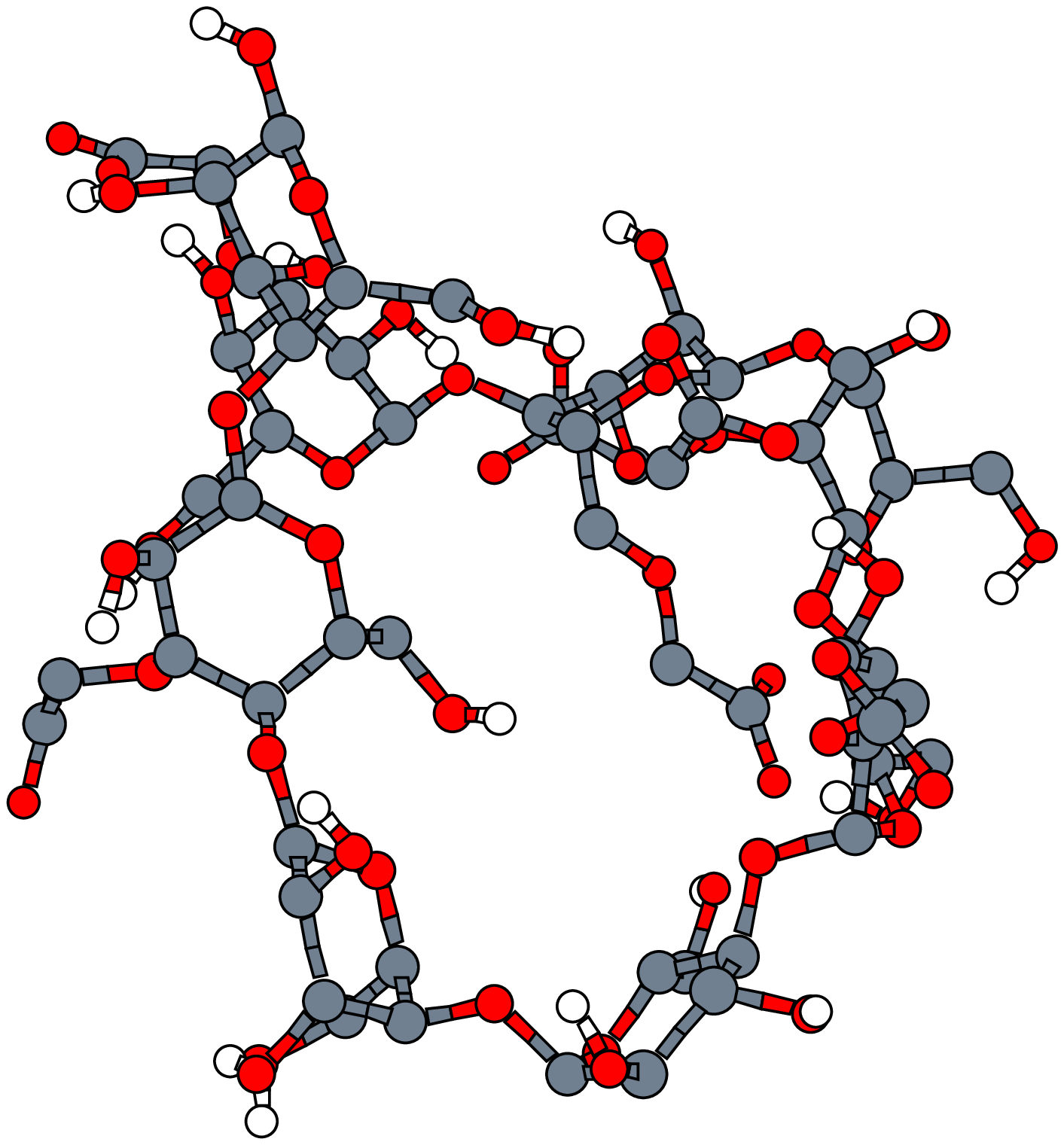}
   \]
   \caption{Snapshots of CMC~I (left) and CMC~II (right).}
   \label{fig:pic-cmc}
 \end{figure}

\begin{figure}
     \Graphics[scale=.6]{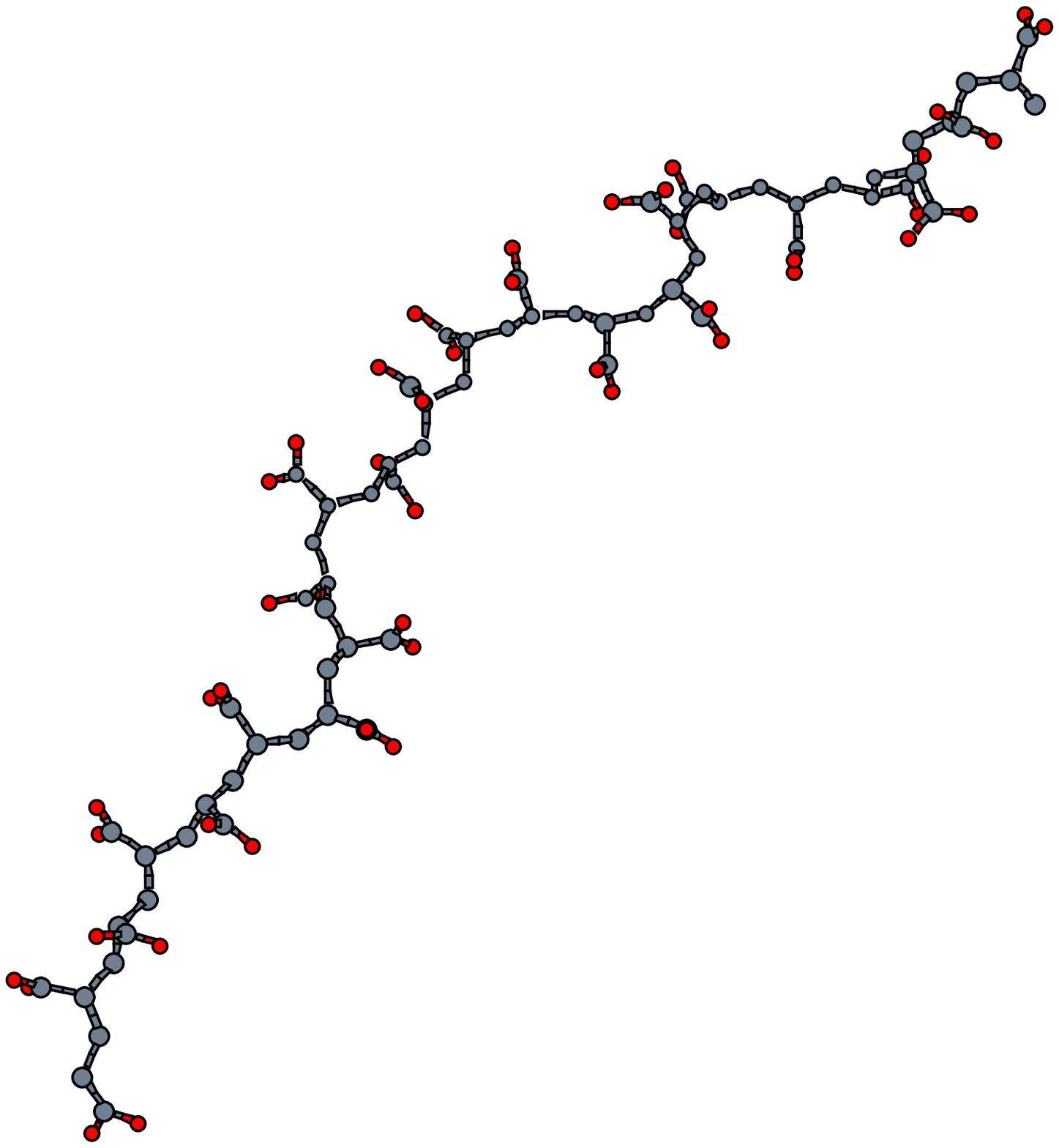}\hspace{-2cm}%
     \Graphics[scale=.6]{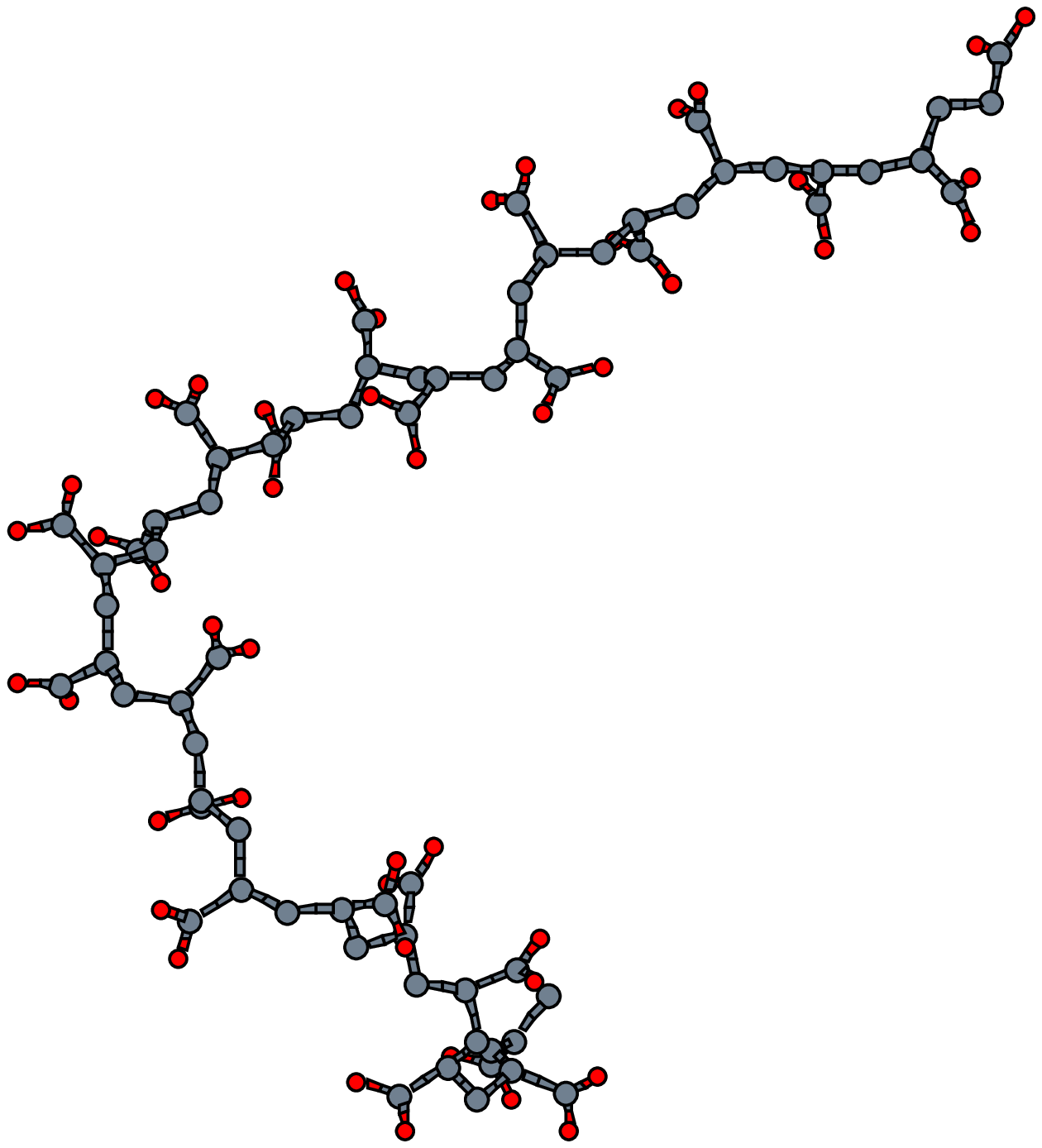}
   \caption{Snapshots of the PAA molecule after $3564\ps$ 
     and $4460\ps$ runtime.}%
   \label{fig:pas1}
 \end{figure}

\begin{figure}
   \[ \Graphics[clip,scale=0.8]{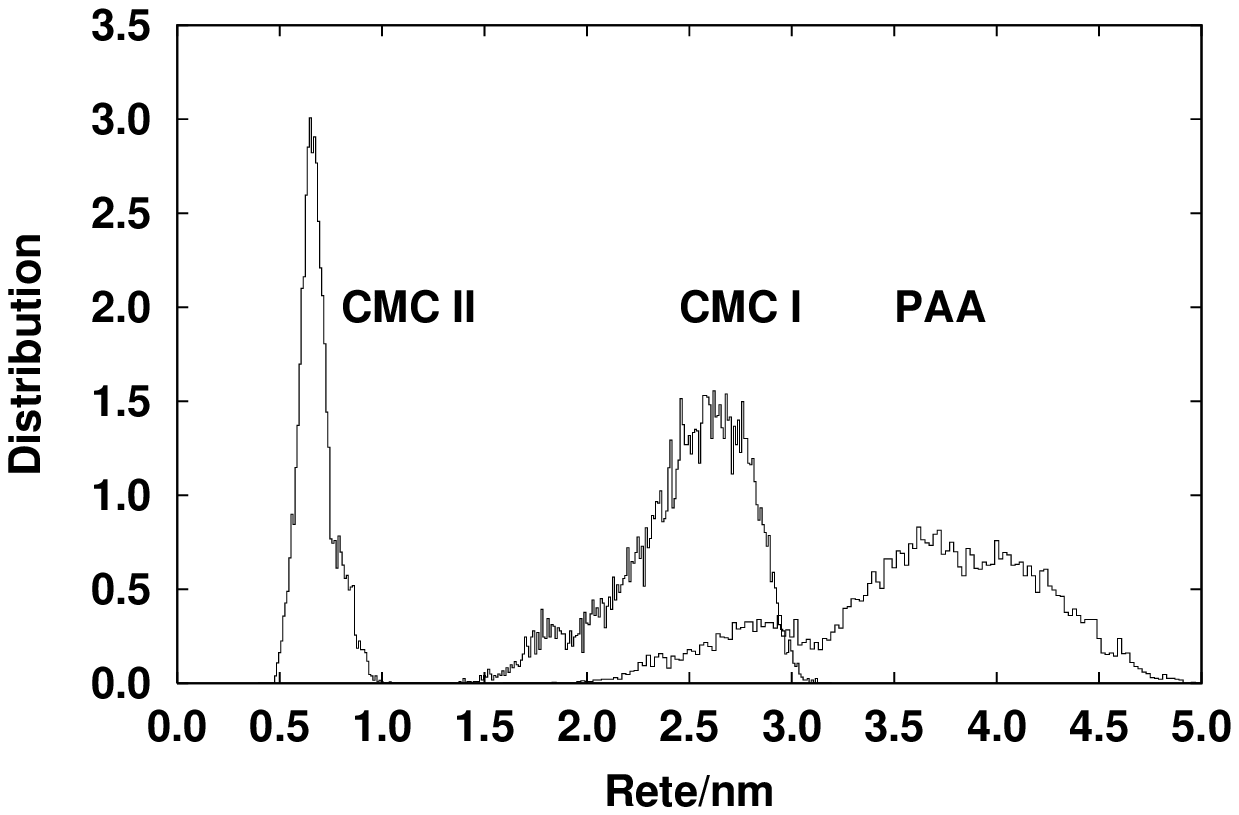}  \]
   \caption{End-to-end distance probabilities  for 
     CMC~I (middle), II (left) and PAA (right).}\label{fig:end2end}
 \end{figure}

 Carboxy methyl cellulose~I (CMC~I) and CMC~II have very different structures:
 CMC~I is an extended, almost straight chain, whereas CMC~II assumes a cyclic
 conformation (Fig.~\ref{fig:pic-cmc}).  Poly(acrylic acid) (PAA) behaves more
 like CMC~I, having mostly a straight conformation with only some bending
 (Fig.~\ref{fig:pas1}).
 These structural features cause the end-to-end distance distributions
 (figure~\ref{fig:end2end}) of CMC~I and CMC~II to be very different. 
 The end-to-end distance $R_\mathrm{ete}$ is calculated between the center of
 mass of the first and last AGU in the chain.  For PAA it is calculated from the
 center of mass of the  terminate repeat units.
 For CMC~I there are two conformers, a major one at $2.6\nm$ and a second
 state at $1.8\nm$.  This second state corresponds to a bent conformation,
 which shows up several times in the simulation.  Its probability is about
 15\Percent\ of the stretched structure and on average this state is kept for
 about $80\ps$ until the oligomer's end-to-end distance lengthens again.  For CMC~II
 the situation is different: The end-to-end distribution is very narrow.  The
 short distance corresponds to a stable ring conformer.  Poly(acrylic acid)
 shows a wider distribution than either CMC, resulting from the larger number
 of torsional degrees of freedom.  The shoulder at $2.8\nm$ is due to the bent
 conformation shown in Fig.~\ref{fig:pas1}.

 The shapes of the molecules are also reflected in the the eigenvalues of the
 gyration tensor (table~\ref{tab:eigenval}).  The stretched conformations of
 CMC~I and PAA, lead to one large and two small eigenvalues, whereas CMC~II
 has more isotropic eigenvalues, with only a factor of two between the largest
 and second largest value.
 \begin{table}
   \[
     \begin{array}{lr@{.}lr@{.}lr@{.}l}
                       & \mcc2{\mbox{CMC~I}}
                                   & \mcc2{\mbox{CMC~II}} 
                                                & \mcc2{\mbox{PAA}} \\
 \avrg*{R_\mathrm{ete}}
                       &  2&5\nm   &  0&70\nm   &  3&6\nm           \\
 \avrg*{R_\mathrm{gyr}}
                             &  0&95\nm  &  0&66\nm   &  1&27\nm         \\
          \avrg*{R_{11}}     &  0&758\nm &  0&245\nm  &  1&346\nm  \\
          \avrg*{R_{22}}     &  0&111\nm &  0&126\nm  &  0&213\nm  \\
          \avrg*{R_{33}}     &  0&047\nm &  0&067\nm  &  0&049\nm  \\
          c_{R_{11} R_{22}}  & -0&86     &  0&14      & -0&76     \\
          c_{R_{11} R_{33}}  & -0&31     & -0&60      & -0&30     \\
          c_{R_{22} R_{33}}  & -0&05     & -0&28      & -0&15     \\
          C_\infty     &  6&7      &  1&0       &  8&3      \\
%          C_\infty>6.0 &\mcl2{76\%} & \mcl2{0\%}   & \mcl2{90\%} \\
    \end{array}
    \]

    \caption{Averages of the end-to-end distance $R_\mathrm{ete}$, the
      gyration   radius 
      $R_\mathrm{gyr}$ and the eigenvalues of the gyration tensor
      ($\avrg{R_{11}}\ge\avrg{R_{22}}\ge\avrg{R_{33}}$) of CMC~I and II  and
      PAA. Additionally, correlation coefficients between the eigenvalues
      ($c_{R_{11} R_{22}}$, etc.) and characteristic ratios $C_\infty$ are given.}
    \label{tab:eigenval}
 \end{table}
 The correlation between the eigenvalues of the gyration tensor follows
 opposite trends (table~\ref{tab:eigenval}): Whereas CMC~I and PAA
 exhibit a negative correlation ($-0.86$ and $-0.76$) between the two
 largest eigenvalues, there is a small positive correlation ($0.14$) for
 CMC~II.  This shows how structural fluctuations happen: CMC~I and PAA
 behave like bending rods -- they shrink in one dimension and
 necessarily grow in another.  In contrast the ring pulsation of CMC~II,
 leads to the two largest eigenvalues growing and shrinking
 simultaneously.
 
 Both the gyration radius and the end-to-end distance have multiple time
 regimes even in the first $80\ps$.  At higher correlation times the
 correlation function's statistics get worse.  
 \begin{figure}
    \[ \Graphics[scale=.8]{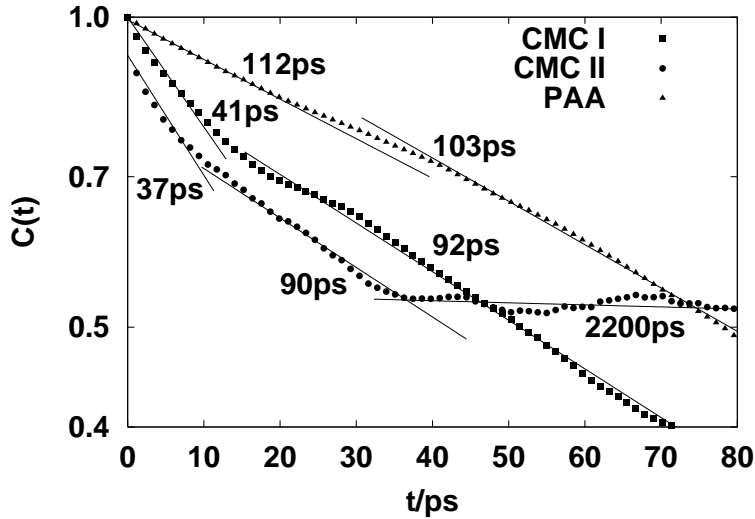} \]
    \caption{Auto correlation functions for the radius of gyration of 
      CMC~I($\blacksquare$), II
      ($\bullet$) and PAA ($\blacktriangle$).  Up to $30\ps$ PAA has the
      slowest decay. (Only every third point shown for
      clarity.)}\label{fig:GyrCf}
 \end{figure}
 The time correlation function of the radius of gyration
 (figure~\ref{fig:GyrCf}) of CMC~I displays two time regimes: on the first
 $15\ps$ a fast one with a time constant of $41\ps$ followed by a slow decay
 with $\tau=92\ps$.  A Kohlrausch-Williams-Watt fit ($f(t)=\exp(-(t/A)^B)$)
 from $0$ to $80\ps$ reveals a total correlation time of $80\ps$ by analytical
 integration~\cite{fit}. Contrary, the CMC~II gyration
 radius decay has four domains: within the first picosecond there is a very
 fast and sharp decay (not resolved in figure~\ref{fig:GyrCf}) followed by a
 decay comparable to the first one for CMC~I with a time constant of $37\ps$.
 The mid-time relaxation is in the same order of magnitude as for CMC~I, but
 there is a vast difference in the long time decay, where CMC~II has a very
 slow mode with a time constant of $\approx2200\ps$.  Averaging the whole
 range from $0$ to $80\ps$ with a Kohlrausch-Williams-Watt fit we calculate a
 total correlation time of $900\ps$.
 
 The decay of the PAA gyration radius is less complex.  It is almost
 monoexponential with a correlation time of $110\ps$.  Compared with
 CMC -- where monomers are larger and intramolecular hydrogen bonding is
 an issue (see section~\ref{sec:HydrogenBonding}) -- PAA lacks any fast
 decay. 
%Thu Nov 23 10:02:25 2000
 One may speculate about the reason for the different relaxation behaviour.
 Poly(acrylic acid) behaves monoexponentially as expected for a generic
 polymer chain in solution.  The relaxation time of $\sim100\ps$ is also
 found for both CMCs at some time scales which supports the idea that it is
 somehow characteristic for a polyelectrolyte of this size in water.  The
 extremely slow long time relaxation (at least 2\ns{}) of CMC~II is also
 easily explained by its rigid globular conformation.  The origin of the
 short-time ($t\lesssim 15\ps$) relaxation behaviour of both CMCs
 (characteristic time $\sim 40\ps$), which is faster than for PAA, is as yet
 not clear.  It is possibly due to the weaker hydration of CMC
 (chapter~\ref{sec:hydrogenSolvent}) which could decrease the short-time local
 friction experienced by the polymer.

 \subsection{Local Chain Properties}
 
\begin{figure}
   \[ \Graphics[scale=0.5]{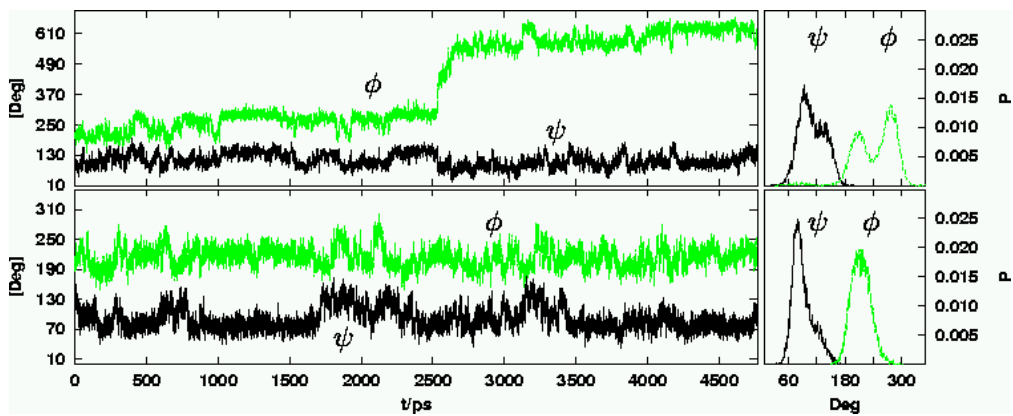} \]
   \caption{Two selected trajectories of ring linking torsions for CMC~I
     ($\phi:$ C4-O4-C1'-C2' and $\psi$: C3-C4-O4-C1').
     The link between rings 1 and 2 is shown at the top and 
     the link from ring 5 to 6 at bottom.  Distribution
     histograms are shown on the right.  Only link 1-2 undergoes a significant
     transition, all other links are fluctuating around with only minor 
     transitions.  They are well represented by the second link shown.}
   \label{fig:tors}
 \end{figure}

 Two selected CMC~I glucosidic torsional angle's $\phi$ (C4-O4-C1'-C2') and
 $\psi$ (C3-C4-O4-C1') trajectories are shown in figure~\ref{fig:tors}.  The
 positions of the maxima differ (for all angles) from results found for
 unsubstituted cellobiose in vacuum~\cite{hardy93} and in
 water~\cite{hardy93b}: For aqueous solutions there are maxima for ($\phi$,
 $\psi$) at ($284$, $63$), ($274$, $193$), ($294$, $242$) and ($50$, $237$)
 degrees (in our definition).  We found most of the linking dihedral angles to
 be unimodally distributed.  Deviation from unimodality depend on the position
 along the chain and the substitution pattern: Torsion $\psi$ is unimodal for
 chain-terminating glucosidic links.  A second $\psi$-state is populated only
 for links connecting repeat unit three with its neighbors.  Occupancies of
 the individual states for CMC~I are given in table~\ref{tab:tors}.  Since
 there are no significant barriers between the states, transitions or, rather,
 oscillations between them occur on a picosecond scale for all torsions.  The
 terminal torsion $\phi$ relaxes faster than the others ($2\ps$ rather than
 $3$--$4\ps$).  Other than this, no significant trends can be found. The
 relaxation time is calculated from the torsional angle auto correlation
 function $\avrg{\Phi(t)\Phi(0)}$ and a subsequent fit and integration.
 
\begin{figure}
   \[ \Graphics[scale=0.5]{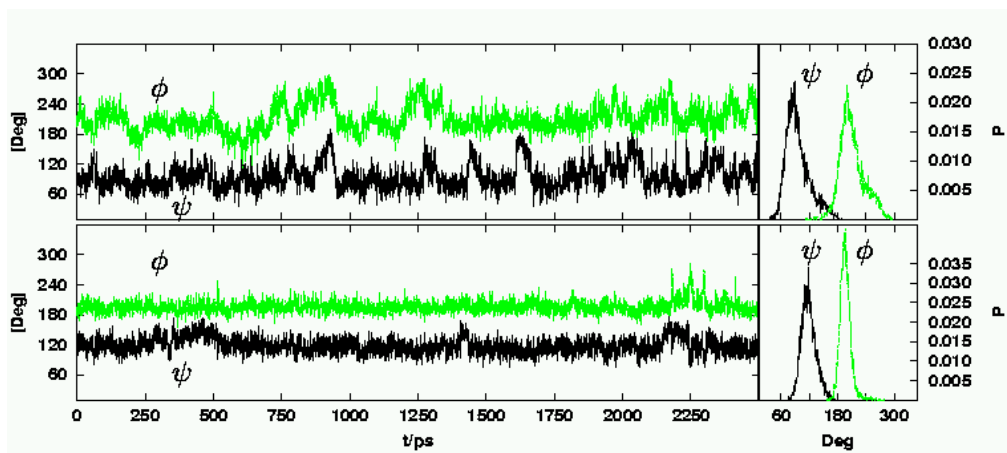} \]
   \caption{Selected trajectories of ring linking torsions for CMC~II
     ($\phi:$ C4-O4-C1'-C2' and $\psi$: C3-C4-O4-C1'), link 1-2 (top),
     link 3-4 (bottom).  The sharp one-state peak for the 3-4 link
     distribution is also characteristic for CMC~II glucosidic links in the
     chain  centre such as links 4-5 or 5-6. Distributions are shown on the
     right.}\label{fig:torsII}
 \end{figure}

 There is hardly any torsional dynamics for CMC~II, which is a
 consequence of the static folded state, with  torsional
 links less mobile than those of CMC~I.  Example torsions are shown in
 figure~\ref{fig:torsII}.

 \begin{table}
     \renewcommand{\rightarrow}{\mbox{--}}
   \[
 \begin{array}{lcccccc}  %time a state is occupied
                  &    \mcc2{\psi\;\unit*{\Percent}}
                                       &&    \mcc3{\phi\;\unit*{\Percent}}\\\cline{2-3}\cline{5-7}
                  & \mcc1{\mathit{0}}
                             &\mcc1{\mathit{1}}
                                       &&\mcc1{\mathit{0}}
                                                   &\mcc1{\mathit{1}}
                                                             &\mcc1{\mathit{2}}\\
\raisebox{1.5ex}[0pt][0pt]{State:}
                 & \mcc1{0^\circ\rightarrow115^\circ}
                             & \mcc1{115^\circ\rightarrow360^\circ}
                                       && \mcc1{240^\circ\rightarrow360^\circ}
                                                   & \mcc1{0^\circ\rightarrow120^\circ}
                                                             &\mcc1{120^\circ\rightarrow240^\circ}\\
 \mbox{Link 1-2}  &  65  &  35         &&  28      &    0    &  72 \\
 \mbox{Link 2-3}  &  59  &  41         &&  57      &    1    &  42 \\
 \mbox{Link 3-4}  &  66  &  34         &&  33      &    0    &  67 \\
 \mbox{Link 4-5}  &  67  &  33         &&  25      &    0    &  75 \\
 \mbox{Link 5-6}  &  85  &  15         &&  11      &    0    &  89 \\
 \mbox{Link 6-7}  &  57  &  43         &&  35      &    0    &  65 \\

    \end{array}\]

     \caption{%
       Inter-ring, glucosidic torsions' occupancy probability (in
       percent) for CMC~I. Each link is characterized by two torsions:
       $\phi$: C4-O4-C1'-C2' and $\psi$:
       C3-C4-O4-C1'. The table summarizes all torsional
       trajectories for CMC~I.  To allow better comparison, we classify the
       torsions into states for each link: The $\psi$-torsion is assigned 
       to two states, the first one covering angles from $0$ to $115$
       degrees, the second one, all remaining conformations.
       The $\phi$-torsion exhibits three possible distinct maxima. One from
       $240$ to $360$, another from $0$ to $120$ and the last 
       from $120$ to $240$.  However, the second state is only
       populated in the link between ring two and three.}
     \label{tab:tors}
   \end{table}
   
  \begin{figure}
    \[ \Graphics[scale=.77]{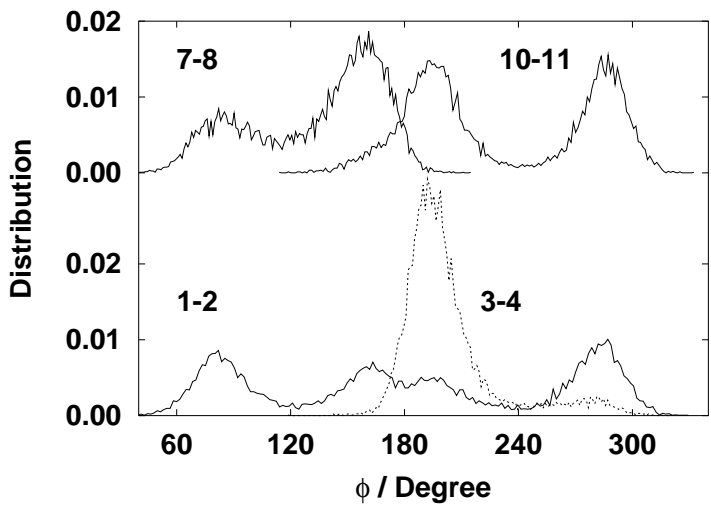} \]
    \caption{Histograms of four representative PAA backbone torsions
      (Torsions between unit seven and eight and ten-eleven are shown on the
      top, torsions between unit one and two, three and four are shown at the
      bottom).  The above two torsions have two maxima, with a clear minimum
      beween them.  The lower histogram shows one almost unimodal torsion and
      a second one with three (or four) maxima.  PAA has two backbone torsions
      per repeat unit.}\label{fig:pastors}
  \end{figure}
  
  Torsions are more complex for PAA (figure~\ref{fig:pastors}), with up to
  three states for each torsion.  The time in which the torsions change
  ($\approx 2$--$5\ps$) is in the same order of magnitude as for CMC~I.  There
  is no apparent correlation between the tacticity of adjacent PAA monomers
  and their backbone torsional statistics.

 \subsection{Hydrogen Bonding}\label{sec:HydrogenBonding}
 
 For biopolymers and modified biopolymers hydrogen bonding (H-bonds) is one of
 the most significant interaction on atomistic scales, both intramolecular
 (conformation) and intermolecular (solvation, aggregation).  Both CMC and PAA
 undergo intensive hydrogen bonding.  In CMC hydroxy groups, carboxylic
 oxygens, and to a smaller extent, ether oxygens are involved.  Poly(acrylic
 acid) can only form H-acceptor bonds via carboxylic groups.
 
 Our definition of a hydrogen bond is geometric: If the two oxygens are within
 a distance of less than $0.29\nm$ with the hydrogen between them, so that the
 angle O-H-O is greater $130^\circ$, a hydrogen bond is assumed.  In our
 polyelectrolyte-water systems, we find both solute-solute, solute-solvent and
 solvent-solvent hydrogen bonding.  The latter is only analyzed in the
 solute's first hydration shell as defined from radial distribution functions
 (below).
 
 The lifetimes $\tau$ of hydrogen bonds may be calculated from
 bond-correlation functions~\cite{mueller-plathe98}: For each possible
 hydrogen-bond we calculate an array, which contains 'true=$1$' in case of a
 bond, otherwise 'false=$0$'. The time-correlation function is calculated on
 this array, integration leads to the lifetime $\tau$.  The integration was
 done by fitting a stretched exponential (Kohlrausch-Williams-Watt)
 $\exp[-(t/\alpha)^\beta]$ in the range $0$ to $80\ps$ and followed
 analytical integration.

 \subsubsection{Solute-Solute Hydrogen Bonding}\label{sec:SoluteSoluteHydr}
 
\begin{figure}
   \[  \Graphics[scale=0.77]{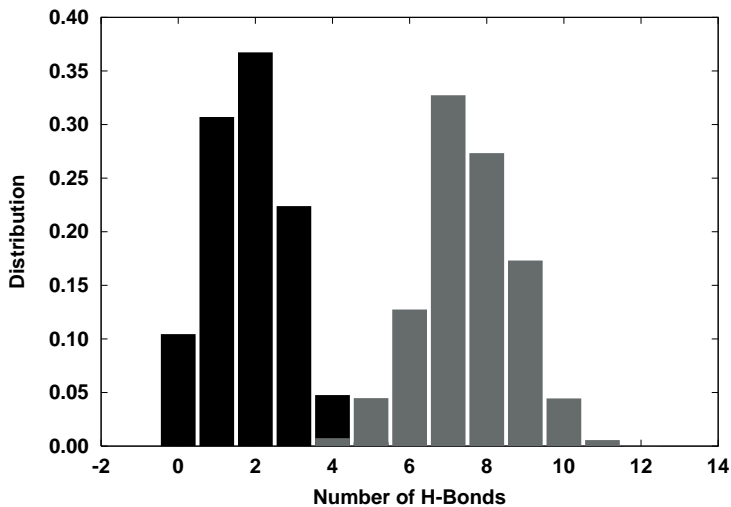}   \]
   \vspace{-2em}
     \caption{Probability distribution of the number of intramolecular
       hydrogen bonds for CMC~I (left) and CMC~II (right).}
       \label{fig:intrahbnds}
  \end{figure}

 The average absolute numbers of intramolecular hydrogen bonds are $1.8$ for
 CMC~I and $7.5$ for CMC~II respectively (see table~\ref{tab:hbonds} and
 figure~\ref{fig:intrahbnds}). The large number of intramolecular hydrogen
 bonds stabilizes the much more compact structure of CMC~II compared to CMC~I.
 
 For CMC, the hydrogen bonds lifetime $\tau$ differs considerably between the
 CMC~I and CMC~II molecule: CMC~I has a bond lifetime of $35\ps$, CMC~II of
 $205\ps$ (table~\ref{tab:hbonds}), indicating that the cyclic compact
 structure of CMC~II is stabilized by long-lived hydrogen bonds.

 \begin{table}
   \[
    \begin{tabular}{l@{\,}ccccc}
           &   & max    & min   & average& $\tau$ \\
       CMC &I  & $5$    & $0$   & $1.8$  &  $35\ps$ \\
       CMC &II & $12$   & $4$   & $7.5$  & $205\ps$ \\
     \end{tabular}
   \]
     \caption{Number of solute-solute hydrogen bonds in one
       conformation:  maximum (max),  minimum (min), average (avrg) and
       lifetime $\tau$ (fit from $t=0\ps$ to $80\ps$).}
     \label{tab:hbonds}
   \end{table}

   \begin{table}
   \[
     \begin{array}{llcccccccc}
                    & n      &   0   &   1   &   2   &   3   &   4   &  5   &   6   &  7   \\
      \mbox{CMC~I}  & P[\%]  &  22   & 78    &   0   &   0   &   0   &  0   &   0   & \mbox-   \\
      \mbox{CMC~II} & P[\%]  &  11   &  18   &  13   &  13   &  18   &  4   & 11    &  12 
      \end{array}
     \]
     \caption{Solute-solute hydrogen bonds for CMC. First row ($n$) 
       contains the distance along the chain between two rings involved
       in a hydrogen bond, the second row ($P$) is probability of
       occurrence.  $n=0$ denotes intra-ring hydrogen bonds.}%
      \label{tab:hbondII}
 \end{table}
 
 Intramolecular hydrogen bonds can be further separated into two
 subsets: \emph{(a)}~intra-ring and \emph{(b)}~inter-ring bonds, which
 connect two distinct glucose moieties and which are presumed to
 influence the overall chain structure of CMC: The ratio of inter- and
 intra-ring bonds for CMC~I is $3:1$.  Lifetimes are shorter for
 intra-ring bonds ($\tau=23\ps$), which demonstrates a lower stability,
 compared to inter-ring bond with $\tau=40\ps$.  Further
 differences can be spotted in the hydrogen bonding atom types:
 intra-ring bonds involve O3-O6 hydrogen bonds ($20\Percent$) and to a
 larger amount O3-COO$^-$ bonds ($70\Percent$).  Bonds between rings involve
 mainly O2-COO$^-$ ($60\Percent$) and O3-COO$^-$ and O6-COO$^-$ ($15\Percent$ each)
 oxygen atoms.
   
 Thus the picture of intramolecular hydrogen bonds in CMC~I is as follows:
 There is a high fraction of long-lived, ring-connecting hydrogen bonds,
 intra-ring bonds are of less importance.  Inter-ring hydrogen bonding happens
 mostly via COO$^-$ groups, which makes \emph{(a)} the interaction energy
 favourable and \emph{(b)} involves a flexible, exocyclic group into binding.
 For this reason, changes in the global chain conformation may be compensated
 by carboxy methyl torsional changes to keep the OH-O alignment in an
 energetically low geometry.  Intra-ring H-bonds, on the other hand, are
 either formed by neutral OH groups, which have a lower energy, or they
 necessitate sharp turns of the carboxy methyl group, which leads to
 distortions.
 
 In CMC~I, inter-ring hydrogen bonds are limited to neighboring rings
 (table~\ref{tab:hbondII}).  This is different for CMC~II, where also more
 distant rings are connected by inter-ring hydrogen bonds: In contrast to
 CMC~I, $90\Percent$ of the inter-ring hydrogen bonds connect rings which are
 not nearest neighbors: on average bonds span $3$--$4$ rings.  These bonds are
 found to be very stable, some of them are intermittently present for an
 overall time of $1.7\ns$, thus living for more than half of the simulation.
 
\begin{figure}
   \[\Graphics[width=\linewidth]{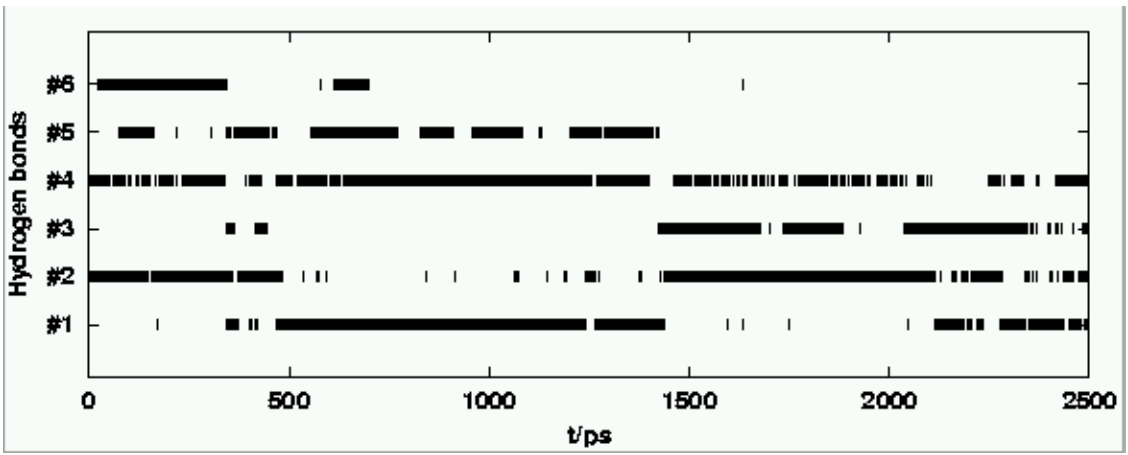}\]
   \caption{Time evolution of six selected hydrogen bonds for
     CMC~II.  For bond pairs (like \#1/\#2 or \#3/\#4) most of the time there is
     only one bond out of a pair intact (Indicated by a bar).  Bond \#1 and \#2
     bridges two glucosidic links, \#3 bridges seven links, \#4 three links and 
     the last two bonds one and  five links, respectively. (Description
     of the H-bonds, subscripts indicate the AGU
     number.  (\#1:O$6_5$-COO${}_7$,
               \#2:O$6_5$-COO${}'_7$,
               \#3:O$3_1$-O$6_8$,
               \#4:O$6_4$-COO${}_7$,
               \#5:O$2_8$-COO${}_7$ and
               \#6:O$6_1$-COO${}_6$)}\label{fig:xorHBII}
 \end{figure}

 Furthermore, in CMC~II there are several pairs of hydrogen bonds (about
 $15$), which bind almost exclusively (see figure~\ref{fig:xorHBII}).  Out of
 these pairs, there is solely one bond present, but alternating with the
 second one.  The presence of one bonds prevents the second bond. Even, if the
 mean lifetime of an individual bond is short, both bonds together form a
 steady connection between two glucose rings.  One example is the bond pair
 O6(ring 3)-COO(ring 7) and O6(ring 5)-COO(ring 7), where the carboxylate
 oxygen flips between the two alcohol oxygens.  Each bond lives $1\ns$ and
 together they span not less than $2\ns$.  However, for $1.7\ns$ only one of
 them is exclusively present.  Thus, for most of the time, one bond excludes
 the other.  This evident for H-bonds formed via a COO$^-$ group, which account
 for every second ($56\Percent$) of all solute-solute bonds.

\begin{figure}
   \[\Graphics[scale=.77]{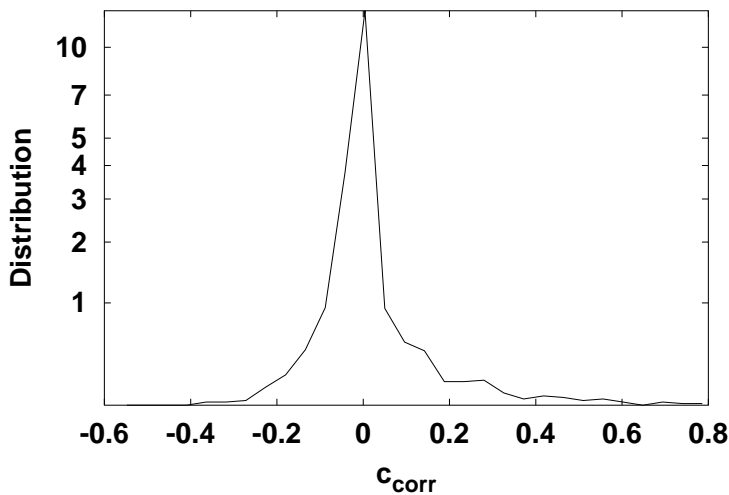}\]
   \caption{H-Bond-H-bond cross correlations for CMC~II.  Distribution of correlation
     coefficients between all intramolecular hydrogen-bonds emerging during the CMC~II
     simulation.  The majority of all hydrogen bonds does not show
     correlations with other bonds.  Only a small part shows (negative and
     positive) correlations. Correlation coefficients are calculated as
     described in section~\ref{sec:SoluteSoluteHydr}.}
   \label{fig:bbcorr}
 \end{figure}

\begin{figure}
   \[\Graphics[width=\linewidth]{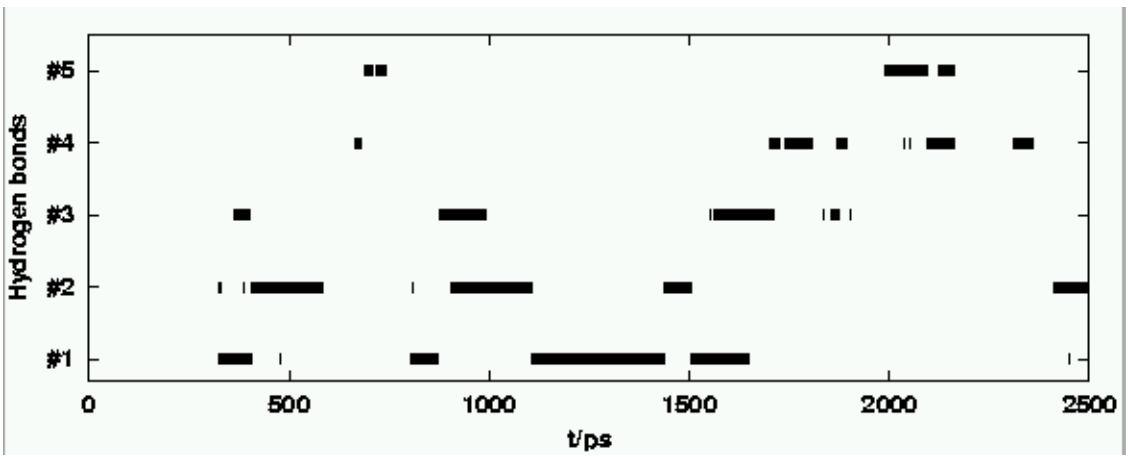}\]
   \caption{Time evolution of four selected hydrogen bonds for
     CMC~I. Hydrogen-bonds \#1 and \#2 are anticorrelated, there is
     only one bond present at any one time.
     (Description of the H-bonds, subscripts indicate the AGU number.
             \#1:O$2_5$-COO${}_6$,
             \#2:O$2_5$-COO${}_6$,
             \#3:O$2_4$-COO$'{}_3$,
             \#4:O$3_7$-COO${}_7$ and
             \#5:O$2_2$-COO${}_3$.)}\label{fig:xorHBI}
 \end{figure}

 A distribution of all bond-bond cross correlation coefficients for CMC~II is
 shown in figure~\ref{fig:bbcorr}.  They are calculated as
 \[
 c_{ij} = %
   \avrg{(h_i(t)-\avrg{h_i})(h_j(t)-\avrg{h_j})}%
     \cdot%
    (\avrg{h_i(t)-\avrg{h_i}}^2\avrg{h_j(t)-\avrg{h_j}}^2)^{-1/2}
 \]
 where $h_i(t)$ is $1$ when hydrogen bond $i$ is present and $0$ otherwise.
 Most hydrogen bonds do not have a significant correlation: $10$
 hydrogen-bond-pairs have a strong negative ($c_{ij}<-0.30$), $45$ bond-pairs
 a positive correlation ($c_{ij}>0.3$).  The anti-correlated bond pairs often
 involve one common oxygen of a COO$^-$-group, which alternates between two
 different hydrogen-donor groups.
 \begin{figure}
   \[\Graphics[scale=0.2]{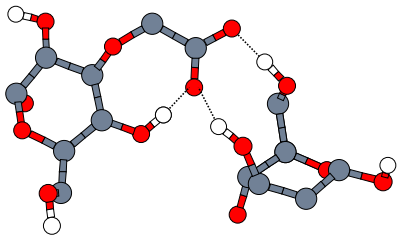}\]
   \caption{Local configuration of solute-solute hydrogen bonds for
     CMC~II. There are two inter-ring H-bonds, which occur only simultaneously
     and one short (oxygen-oxygen distance: $2.4\nm$) intra-ring bond which
     forms an eight-membered-ring.  The two inter-ring bond occur only simultaneously.}
   \label{fig:embeddcoo}
 \end{figure}
 High correlations ($c_{ij}>0.6$) occur particularly, if the compact
 configuration of the CMC~II oligomer allows for some bonds to form
 simultaneously, like H-bonds involving O2H and O6H donor groups of one, and a
 single carboxylic group of a second ring.  The COO$^-$ group is embedded between
 the two H-donors (figure~\ref{fig:embeddcoo}).  The effect is very strong for
 the flexible COO$^-$ group at rings 7 and 8, but is still noticable for other
 acceptors.  Examples of hydrogen bond trajectories (Fig~\ref{fig:xorHBI})
 illustrate the cases of (anti) correlated H-bonds.
 
 The cross correlation coefficient $c$
 \begin{equation}
   c = 
       \frac{\avrg{(n-\avrg{n}) (r - \avrg{R_\mathrm{ete}}) } }
            {\sqrt{\avrg{(n-\avrg{n})^2} \avrg{(R_\mathrm{ete} - \avrg{R_\mathrm{ete}})^2}}}.
 \label{eq:1}
 \end{equation}
 shows some coincidence between the number of solute-solute hydrogen-bonds $n$
 and the polyelectrolyte's end-to-end distance $R_\mathrm{ete}$ for CMC~I.
 Whereas there is hardly any correlation of extension and the total number of
 intra-CMC hydrogen-bonds ($c=-0.05$), it is slightly more pronounced
 ($c=-0.16$) for the number of intra-ring hydrogen bonds versus the end-to-end
 distance: If the molecular extension shrinks, the number of intramolecular
 hydrogen bonds increases.  The compact polymer structure favours more
 contacts.  Similar correlations for CMC~II would not be very meaningful,
 because of the rigid structure which fixes the end-to-end distance.

 \subsubsection{Hydrogen Bonds to Solvent}\label{sec:hydrogenSolvent}
 
\begin{figure}
   \[ \Graphics[scale=0.7]{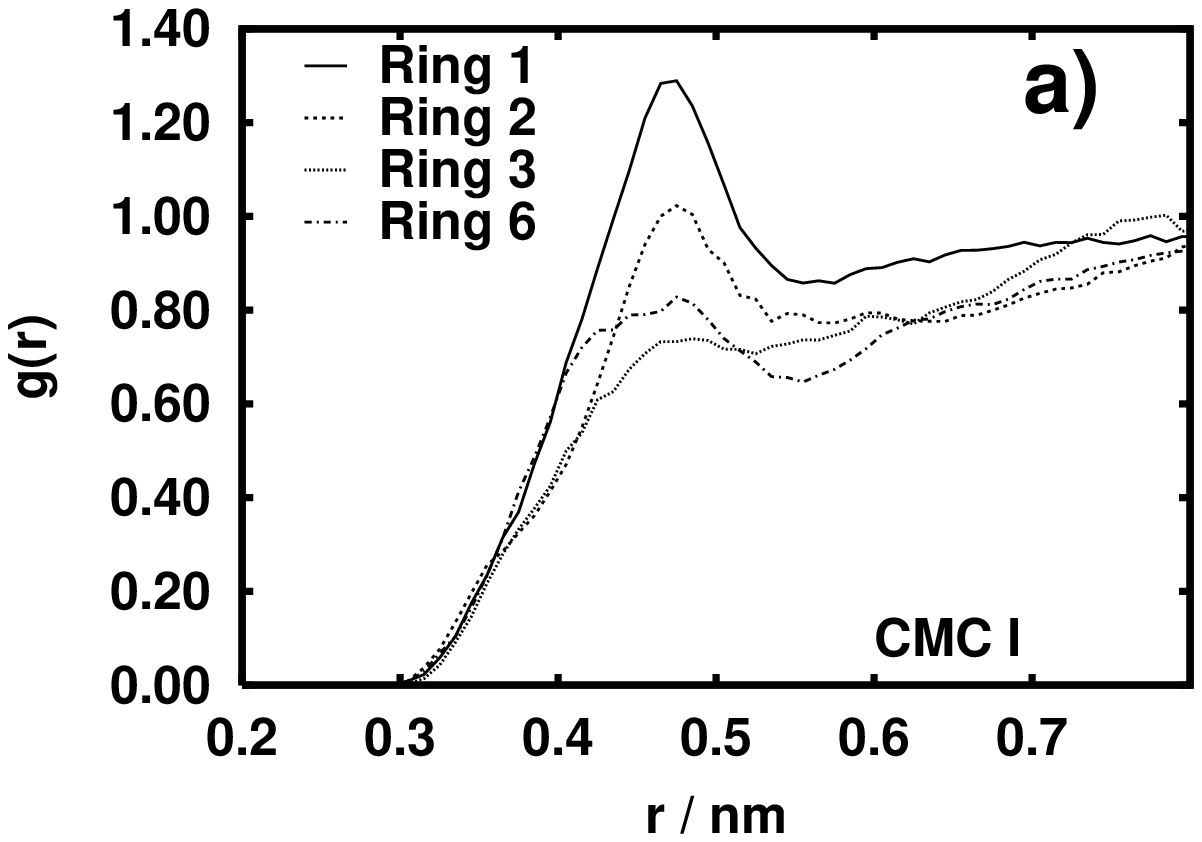}  \hspace{-1.8cm} \Graphics[scale=0.7]{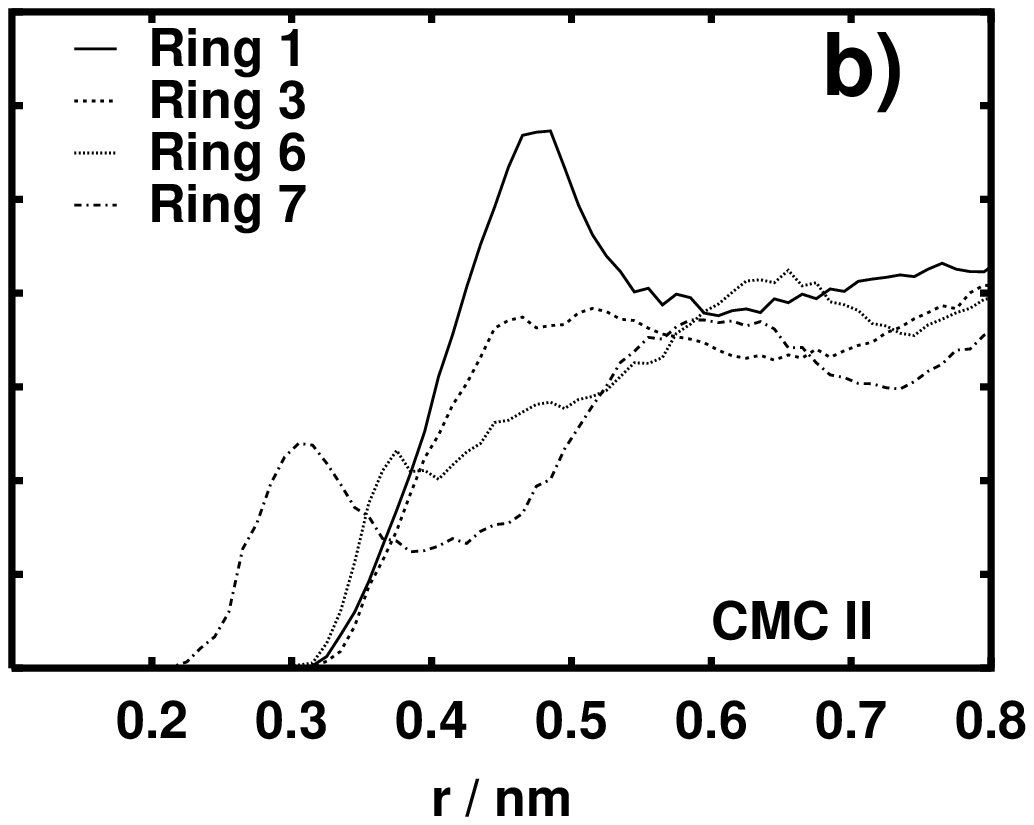}  \]
   \[ \Graphics[scale=0.7]{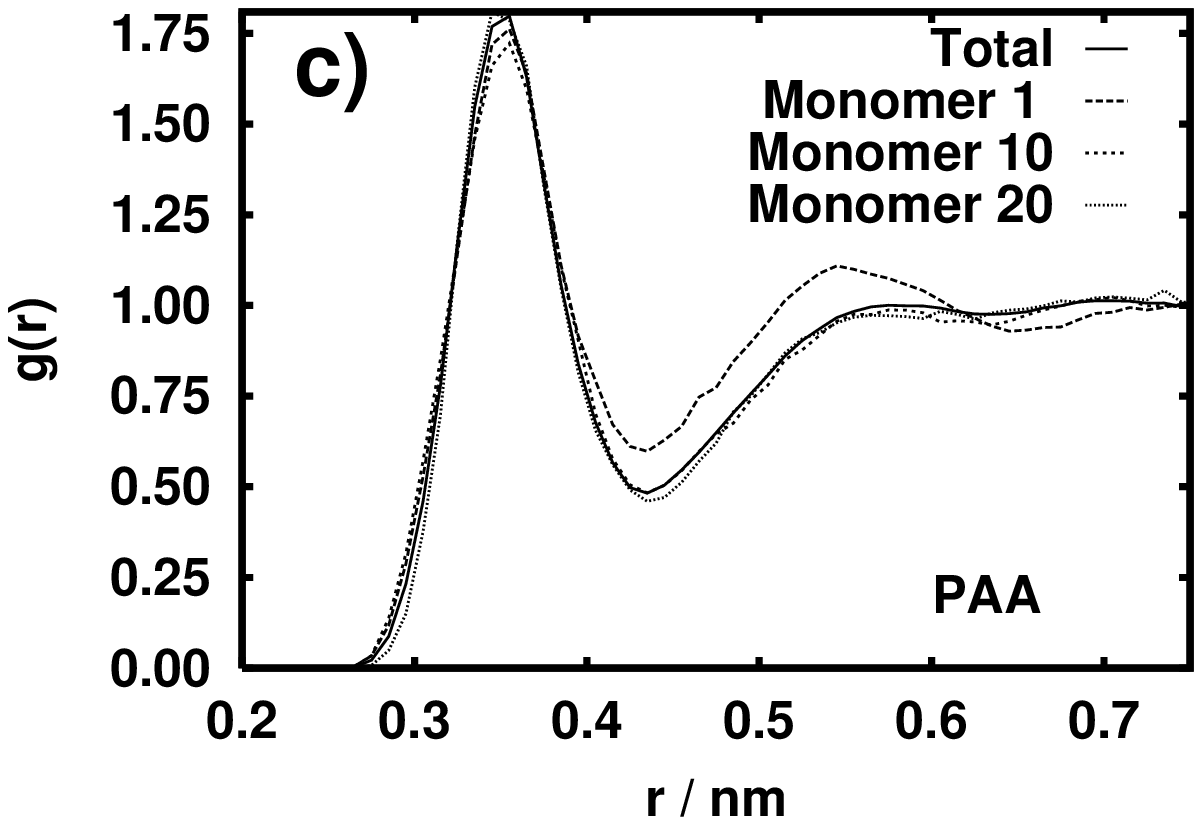} \]
      \vspace{-2\baselineskip}
      \caption{Pair distribution functions $g(r)$ of selected CMC glucose rings
        and PAA monomer units (center of mass) with water oxygens.  \emph{a)} CMC~I,
        \emph{b)} CMC~II, c) PAA.  For CMC~I one identifies a first hydration shell
        with on average $15$ next neighbors to the first minimum.  The first
        shell of CMC~II is more irregular, with hydration numbers ranging from
        $10$ to $18$.  PAA has the most regular and smooth hydration shell.
        As a general result, end monomers have more water molecules in their
        neighborhood than inner monomers.}\label{fig:com-ow}
 \end{figure}

 A general picture of the solvent distribution around CMC is given by
 radial distribution functions (RDF) between the center of mass (defined as in
 section~\ref{sec:CompDeta}) of the glucose unit and water oxygens
 (figure~\ref{fig:com-ow}).
 The differences in the peak heights can be described and explained as
 follows: The outermost glucose rings exhibit the best defined hydration
 shell.  Environments of inner rings are more perturbed, either by neighboring
 rings or by attached carboxylic groups.  This is evident for the doubly
 substituted ring 3 in CMC~I, whose first solvation peak disappears entirely.
 The CMC~II radial distributions for rings 1 and 3 show multiple peaks with
 the first maximum at approximately $0.45\nm$.  Only ring 7 shows two
 different peaks centered at $0.3\nm$ and $0.6\nm$, where the first peak at
 $0.3\nm$ is caused by the solvation of the C6 carboxy methyl group.  Thus
 from the appearance of the RDFs, the CMC~I solvation shell seems to be more
 intact than that at CMC~II.  This is the consequence of the open-chain
 configuration of CMC~I as opposed to the compact structure of CMC~II, which
 excludes solvent molecules from the vicinity of different monomers to a
 different extent.
 
 As PAA has a simple configuration, the solvation shell is more uniform:
 the chain ends attract more water than inner segments, but, apart from
 that, the center of mass-water RDF is the same with the first peak at
 $0.35\nm$.
 
\begin{figure}
   \[ \Graphics[scale=.8]{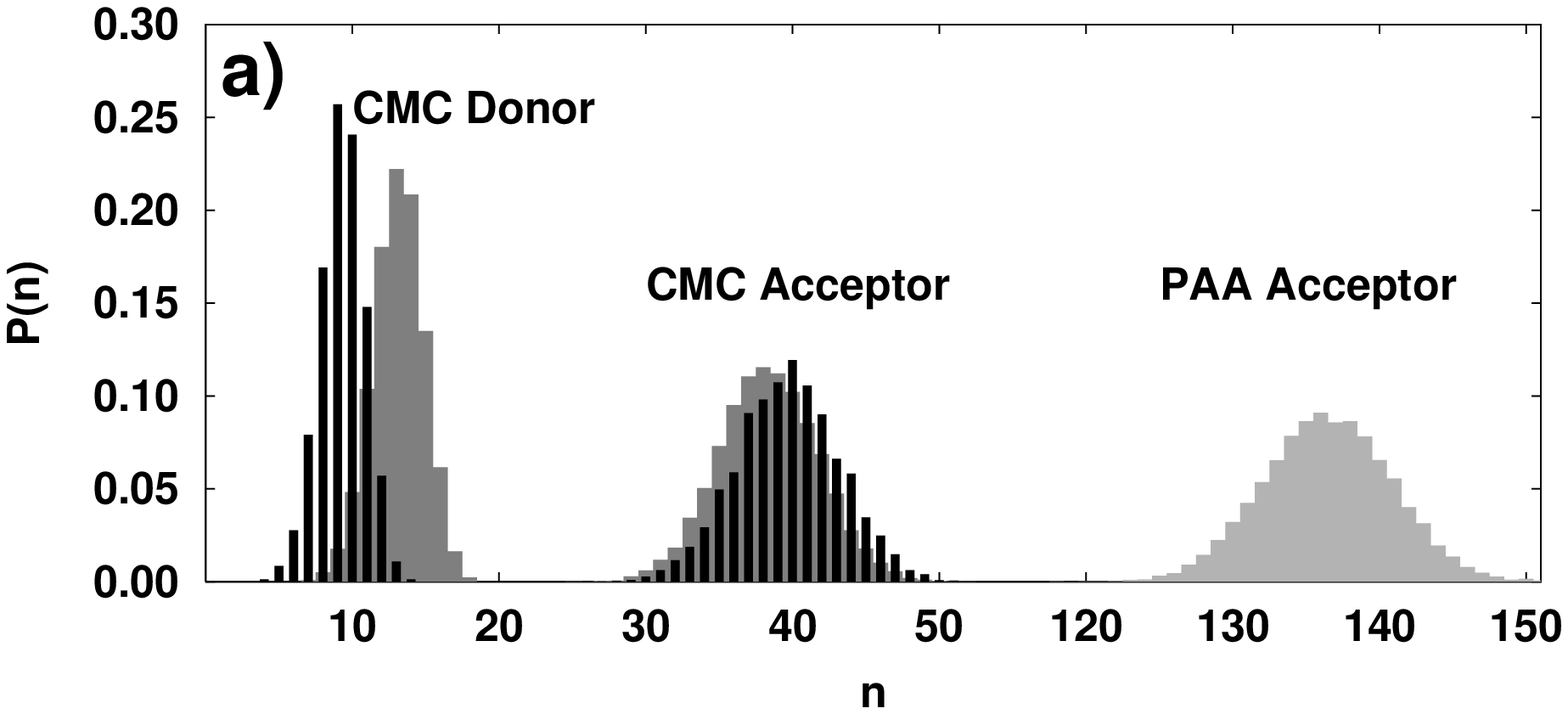}  \]
   \[ \Graphics[scale=.8]{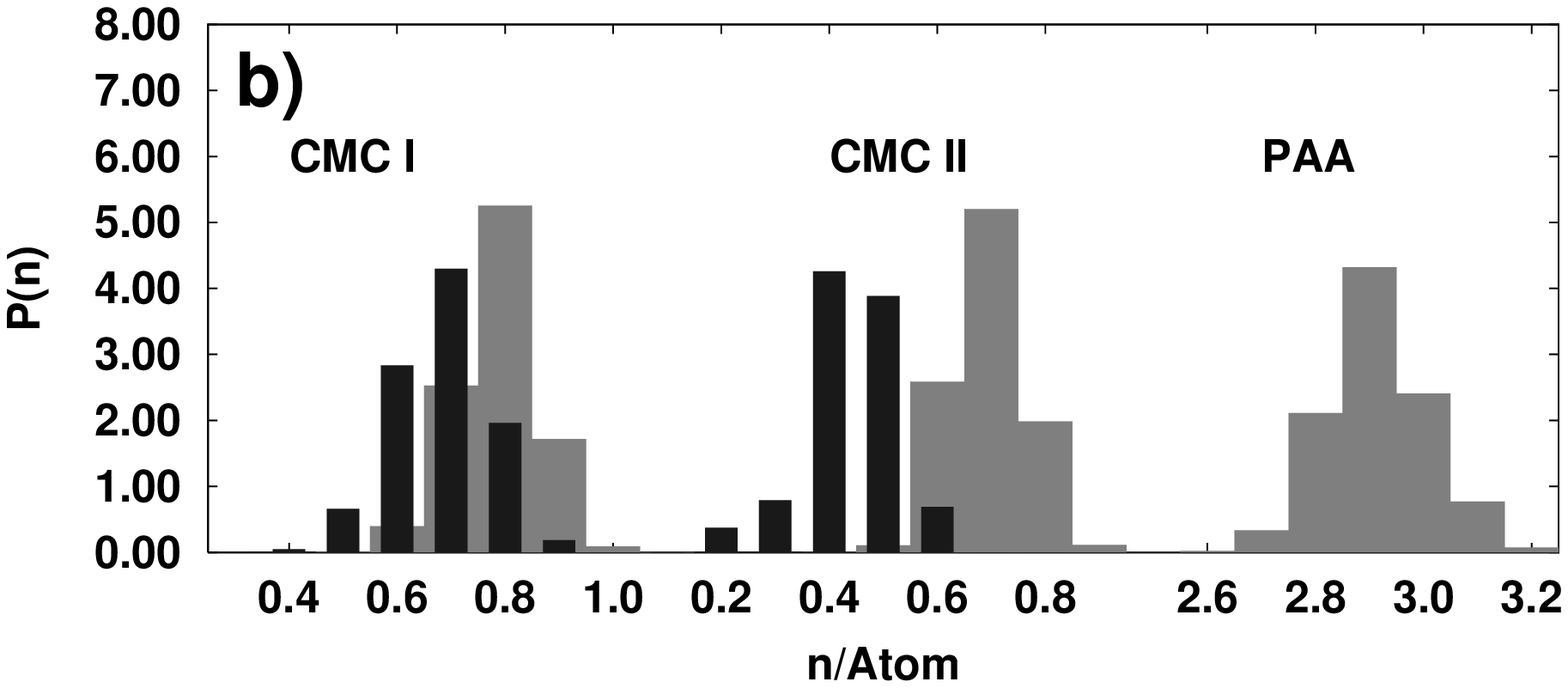} \]
   \caption{%[]{
     \emph{a)} Total number of donor and acceptor hydrogen bonds
     between water and CMC~I (grey) and CMC~II (black) and PAA.  \emph{b)}
     Histogram of the number of solute-solvent hydrogen bonds for CMC~I, CMC~II
     and PAA (from left to right).  Normalized with respect to the appropriate
     number of binding sites of the polyelectrolytes (all polymer-oxygens as
     acceptor, all hydroxy-groups as donor to water).  Acceptor bonds are
      grey, donor bonds are dark.}\label{fig:hhist}
 \end{figure}

 On average the CMC~I oligomer has $13$ donor and $38$ acceptor hydrogen bonds
 with water, CMC~II $9$ donor and $35$ acceptor bonds
 (table~\ref{tab:interHbonds}).  The difference is in line with the result for
 intramolecular bonds: The CMC~II molecule, which heavily bonds with itself,
 has fewer free sites for hydrophilic interaction with the solvent.
 Poly(acrylic acid), with no hydrogen bond donors of its own, has on average
 no less than $136$ bonds with water.  As an estimate, this corresponds to
 water molecules with a total mass of $2300$\,amu, which is $1.5$ times the
 polymer's own mass.  The mass of associated water is smaller for CMC: About
 $60\Percent$ of the polymer's mass for CMC~I and only $50\Percent$ for
 CMC~II.  The detailed distributions of H-Bonds are shown in
 figure~\ref{fig:hhist}.  They are approximately Gaussian in all cases and
 they reflect the trends already seen in the averages.

 \begin{table}
   \[\begin{array}{l*{3}{llc}}
                          & \mcl2{\mbox{CMC~I}}   
                                            && \mcl2{\mbox{CMC~II}}
                                                               && \mcl2{\mbox{PAA}}   \\
   \mbox{Donor-OH}        &  13  & (0.72)   && 9   &(0.45)     &&       &           \\
   \mbox{Acceptor COO$^-$}&  26  & (2.60)   && 23  &(1.92)     &&  136  & (2.9)     \\
   \mbox{Acceptor OH}     &  10  & (0.56)   && 11  &(0.55)     &&       &           \\
   \mbox{Acceptor ether-O}&  1.5 & (0.10)   && 1.3 &(0.01)     &&       &           \\
    \end{array}\]
   \caption{Average number of hydrogen bonds to solvent, grouped according 
     to solute binding sites.  In parentheses: numbers with respect
     to hydroxy, ether and carboxylic oxygen sites of the
     polyelectrolytes.  For the two CMC oligomers, there is
     hardly any difference in the number of OH-acceptor H-bonds per site.
     In contrast the number of CMC~I donor-OH and acceptor COO$^-$ bonds
     per site is lower  than for CMC~II.  From this follows that the
     intrapolymer H-bonding happens in many cases trough --OH-COO$^-$ bonds.}
   \label{tab:interHbonds}
 \end{table}

 In order to compare the hydration of the different molecules on an equal
 basis, one may normalize the number of H-bonds by the mass of the polymer.
 This yields the H-bonds per weight, which should in some way be correlated
 with the specific enthalpy of solvation.  One finds $0.036$\,bonds/amu for
 CMC~I, $0.029$\,bonds/amu for CMC~II and more than twice as many for PAA:
 $0.083$\,bonds/amu.  Thus there seems to be a competition between inter- and
 intra-molecular hydrogen bonds for carboxy methyl cellulose, which is
 dominated by the intramolecular bonds in case of CMC~II and to a lesser
 extent for CMC~I.  In PAA, there is no competition from intramolecular
 hydrogen-bonds.  In addition, one can reasonably expect that hydrogen-bonds
 to COO$^-$ with its negative charge are energetically more favorable than to
 --OH or --O--.  Hence, PAA should have a more negative enthalpy of solvation
 than CMC.  We are, however, only aware of experimental data for PAA to
 compare with~\cite{klein79}.  Klein et al. used solution calorimetric
 measurements to determine the solution enthalphy of about
 \mbox{$25\unit{kJ}$/(mol repeat unit)} ($0.27\unit{kJ/g}$) for sodium-PAA,
 this value corresponds to $575\unit{kJ/mol}$ for  the PAA 23-mer.

 \subsection{Structure of the solvation shell}

 The arrangement of water molecules in the vicinity of the polymer
 chains has been investigated. To this end, all water molecules within
 $0.6\nm$ of any polymer atom were sorted into clusters. A water molecule
 was taken to be part of a cluster if it formed a hydrogen bond to
 any other water molecule of that cluster. Hence, the assignment
 of molecules to clusters depends on our geometrical definition of
 a hydrogen bond (Sect.~\ref{sec:HydrogenBonding}).
 
\begin{figure}
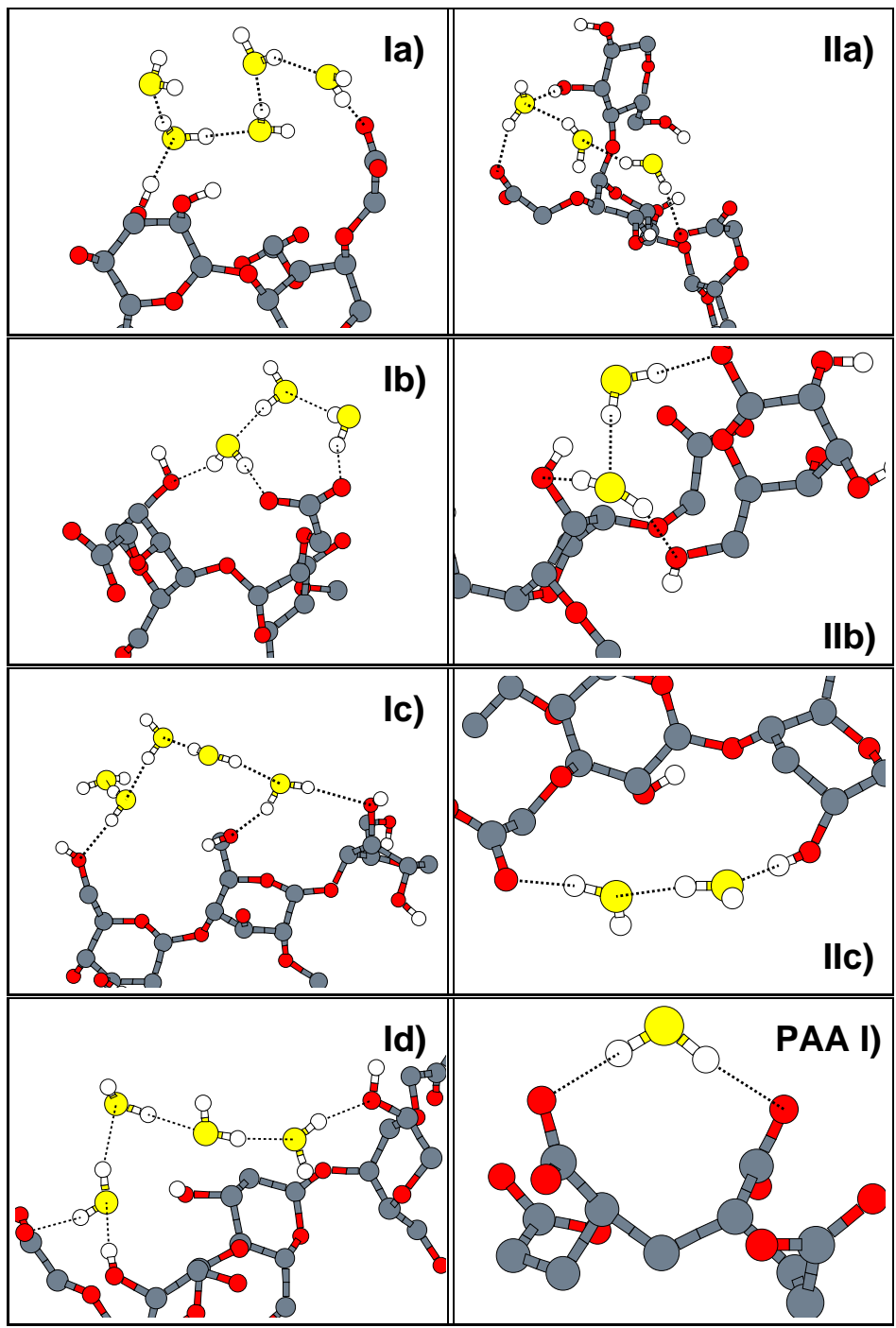

   \[\Graphics{clusterI}\]
   \vspace{0cm}
   \caption{Examples of water cluster in the vicinity of CMC~I, II and
     PAA.  Dotted lines denote hydrogen bonds.}
   \label{fig:clusterI}
 \end{figure}

\begin{figure}
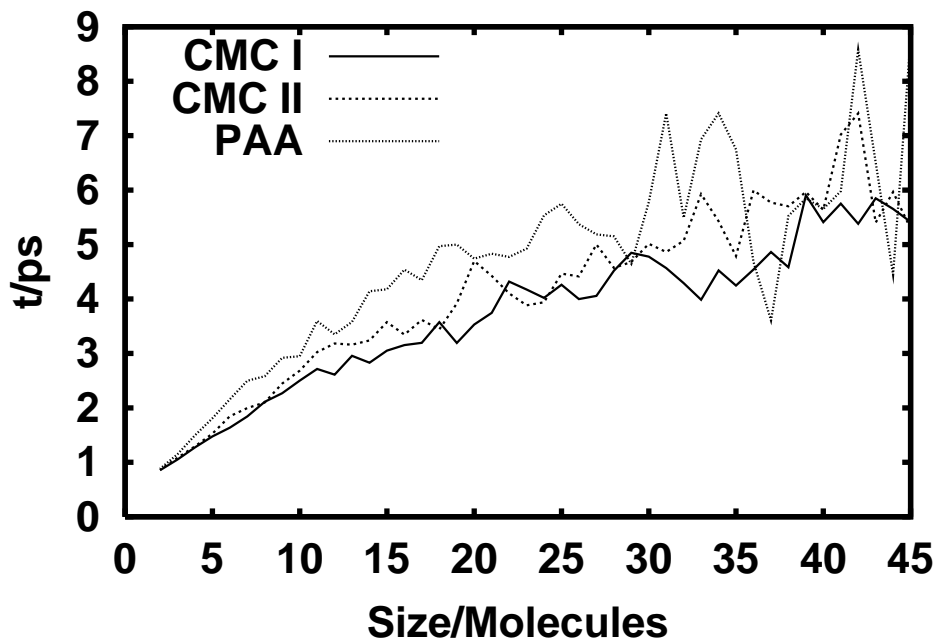

   \[\Graphics{MSizeVsLife}\]
   \caption{Average  lifetimes of water clusters in the vicinity of the
     polyelectrolytes as a function of size.  Two pathways are responsible for
     the disappearance of a cluster: Either it is overtaken by a larger one
     and inherits his identity, or it diminishes by loosing all of its
     molecules.}\label{ClusterLifeTime}
 \end{figure}

 For the three polymers, examples of water clusters in the immediate vicinity
 of the solute are shown in Figure~\ref{fig:clusterI}. For both CMCs, such
 clusters consist of $2$--$5$ molecules and are attached to the solute via
 $2$--$3$ hydrogen bonds (in addition to individual water molecules
 hydrogen-bonded to CMC which are not shown). Depending on their size, the
 clusters can connect two (CMC~I: a, b; CMC~II: a-c) or more (CMC~I: c,d)
 cellulose rings.  Poly(acrylate) is quite different due to the shorter
 distances between repeat units. Typically, one finds single water molecules,
 that are not part of water clusters and that bridge neighboring carboxylate
 groups (Fig~\ref{fig:clusterI}, PAA I).  The clusters are, however, fragile
 arrangements undergoing rapid exchange (order of picoseconds,
 Fig~\ref{ClusterLifeTime}) with other water molecules and they do provide
 little if any stabilisation to particular CMC conformations.  Only a small
 difference is found between both CMC oligomers
 (figure~\ref{ClusterLifeTime}), whereas clusters in the hydration sphere of
 PAA tend to live longer.  This could be due to the high concentration of
 negative charges on the PAA backbone, which slow down the general motion of
 hydration water by their strong electrostatic field.

%%%%%%%%%%%%%%%%%%%%%%%%%%%%%%%%%%%%%%%%%%%%%%%%%%%%%%%%%%%%%%%%%%%%%%
%%
%%
%%
%%
%%
%%%%%%%%%%%%%%%%%%%%%%%%%%%%%%%%%%%%%%%%%%%%%%%%%%%%%%%%%%%%%%%%%%%%%%

  \subsection{Counterions}
  
  The distribution of counterions around the solute molecules has also been
  investigated. The solutes are fragments of polyelectrolytes, and since the
  work of Manning~\cite{manning69} the distribution of counterions in the
  field of polyelectrolytes has been of considerable theoretical interest
  (see, e.g., ref.~\cite{barrat96} and references therein).  Because of the
  small size of the charged solutes investigated here and the small number of
  counterions, this study has to be limited to the molecular neighborhood of
  the solutes.
  
\begin{figure}
   \[
   \Graphics[scale=0.65]{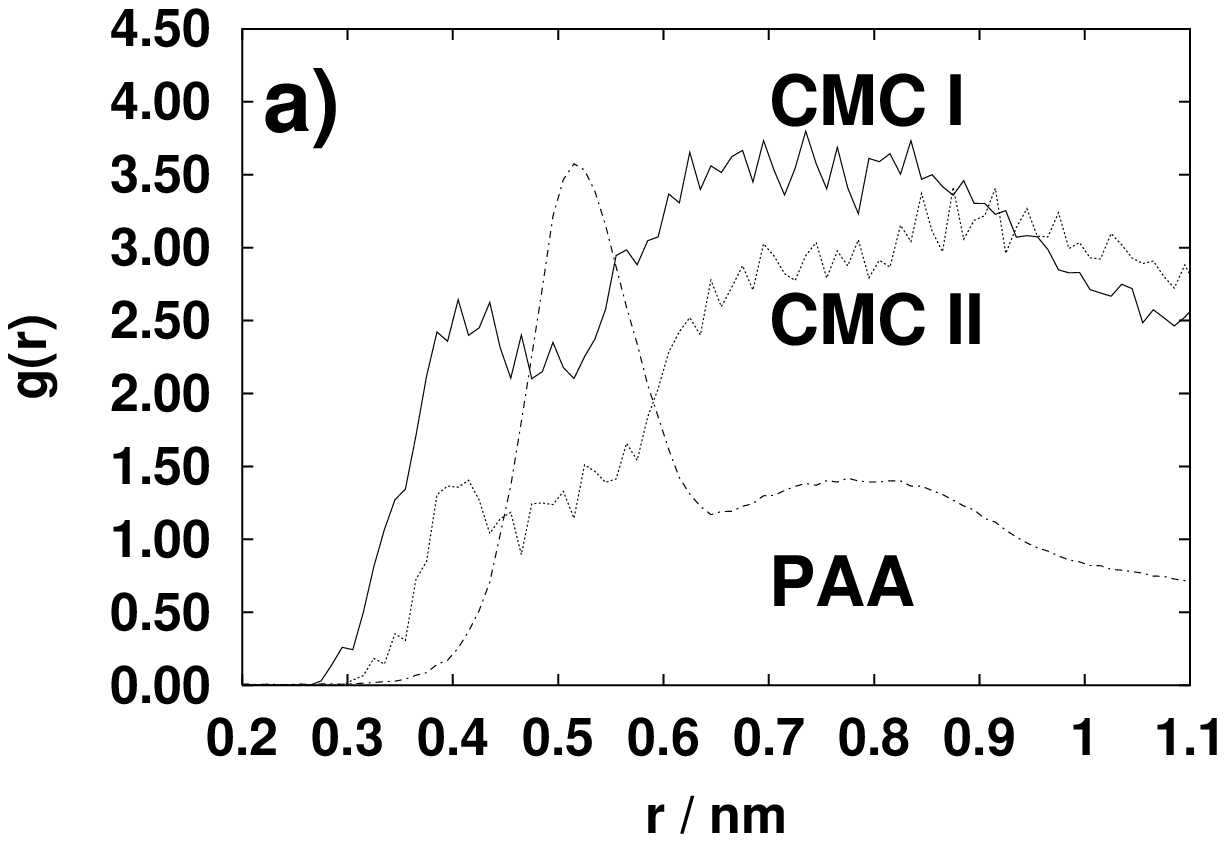}
   \Graphics[scale=0.65]{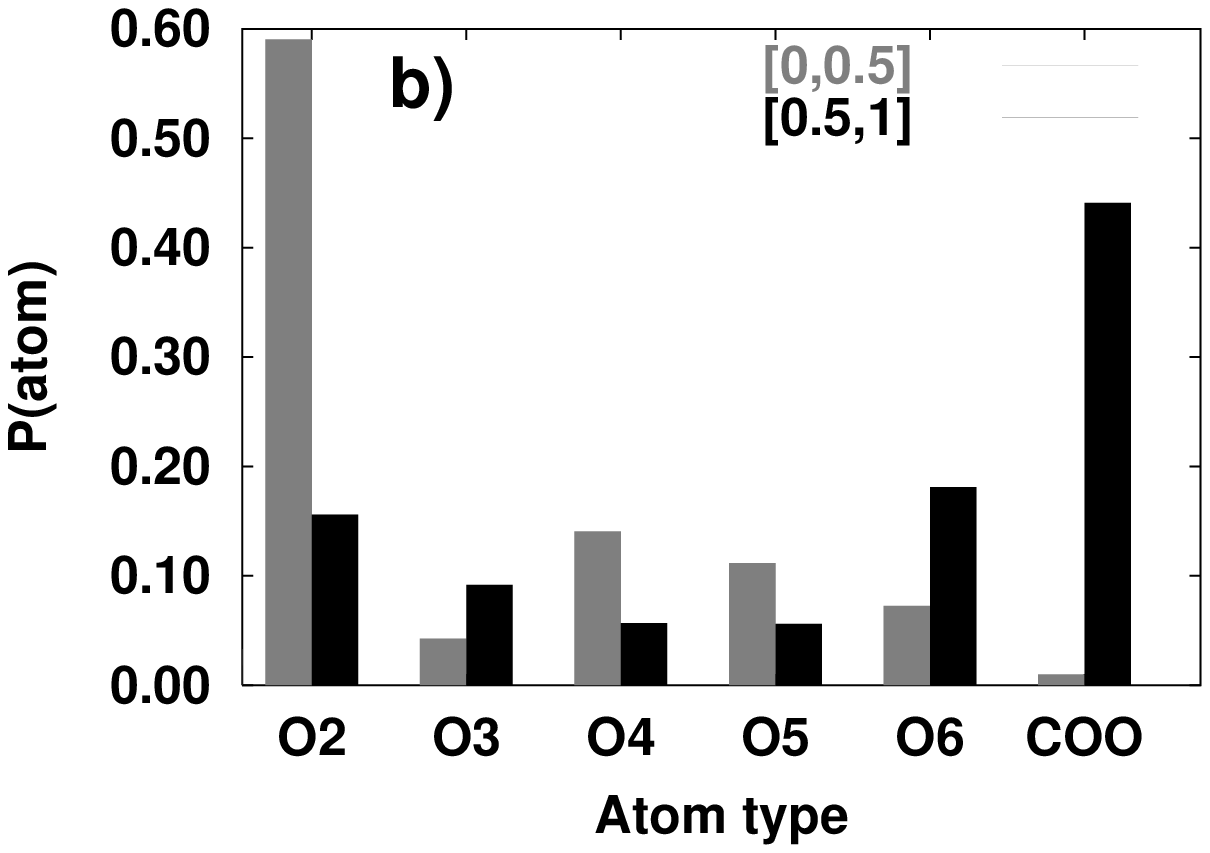} 
   \]
   \caption{
     \emph{a)}~Monomer(center of mass)-sodium radial distribution.  Both CMC
     molecules show the same shape RDF, but CMC~II has smaller peak heights.
     All radial distributions reach unity at about $r=1.9\nm$.  \emph{b)} The
     figure shows the preference of docking sites for sodium ions approaching
     CMC~I.  The histogram is divided into two parts. The first (grey)
     includes ions close to the chain (distance sodium-ring-com lower
     $0.5\nm$), the second one (black) is for all ions in the distance of the second
     peak in the pair distribution function ($\in[0.5\nm, 1.0\nm]$).}\label{fig:comNa}
 \end{figure}

  Radial distribution functions (RDFs) between the centres of mass of the
  repeat units (for definition, see Sect.~\ref{sec:CompDeta}) and the Na$^+$
  are summarised in Figure~\ref{fig:comNa}\,a.  Both CMCs are similar in
  shape, but different in intensity: They have a close peak at about $0.4\nm$
  and a broad second peak centered at about $0.7\nm$.
  Figure~\ref{fig:comNa}\,b shows that the first peak ($r<0.5\nm$) is
  dominated by Na$^+$ approaching the O2 oxygen of the cellulose moiety,
  whereas the second peak ($0.5$--$1\nm$) arises from Na$^+$ ions near the
  carboxylic groups which, in CMC, are more remote from the centre of mass.
  Both the (dynamic) flexibility of the carboxylate and the (static) different
  substitution patterns on different AGUs account for the large width of the
  second peak. CMC II has a lower intensity, especially at short distance.
  This arises from its more compact structure which prevents counterions from
  approaching it freely from all sides.  This behaviour is similar to what was
  already seen for the solute-solvent H-bonds.
  
  The RDF of poly(acrylate) is better defined due to the larger number of
  counterion-monomer pairs. It shows a relatively sharp first maximum at about
  $0.5\nm$ caused by Na$^+$ ions near the carboxylate and a broader second maximum
  ($\approx0.8\nm$) possibly due to solvent-separated ion pairs.
  
  It is interesting to note that on average the CMC oligomer keeps much fewer
  counterions ($\approx0.41$) in its electrostatic vicinity (distance to any
  atom of the polymer lower than the electrostatic screening length of the
  solvent $<0.64\nm$) than the PAA oligomer ($\approx13$), in spite of both
  being of similar molecular weight.  This is caused by the difference in
  charge.

%%%%%%%%%%%%%%%%%%%%%%%%%%%%%%%%%%%%%%%%%%%%%%%%%%%%%%%%%%%%%%%%%%%%%%
%%
%%
%%
%%
%%
%%%%%%%%%%%%%%%%%%%%%%%%%%%%%%%%%%%%%%%%%%%%%%%%%%%%%%%%%%%%%%%%%%%%%%

 \section{Conclusions}
  
  Even though poly(acrylate) and carboxy methyl cellulose both are
  water-soluble polyelectrolytes, their behaviour in water and towards water
  differs markedly.  This is due to the different charge density as well as to
  the different type and quality of hydrogen bonds that either forms with
  water.  In PAA, there is one strong hydrogen bond with the deprotonated
  carboxylate acting as an acceptor.  In CMC, the smaller density of
  carboxylates is only partly set off by the possibility of forming both donor
  and acceptor hydrogen bonds to the alcoholic OH groups of the cellulose.
  (Hydrogen bonds to the ether oxygens are irrelevant.) Taken per molecular
  weight of the polymer, it seems safe to say that PAA forms at least twice as
  many hydrogen bonds as CMC and that they are of larger binding energy
  (charge-dipole, rather than, dipole-dipole). Based on this argument, the
  solvation of PAA in water should be more exothermic than that of CMC.
  Unfortunately, no measurements appear to be available for comparison.
  
  The comparison of the two CMC oligomers shows that the particular
  carboxy methylation pattern has an immense influence on the local structure
  in solution. The two assume entirely different conformations: CMC~I is
  stretched and flexible, whereas CMC~II remains in a rigid cyclic
  conformation.  While it cannot be ruled out completely that one molecule may
  not have reached equilibrium and may still be stuck in a local minimum, this
  is unlikely in view of the long simulation times of several nanoseconds. If
  there are metastable conformations they have to be long-lived.  We are
  therefore left to conclude that industrial CMC with its statistical
  substitution of OH groups, behaves locally very diversely.  As a consequence
  of its globular structure, CMC~II shows more intramolecular hydrogen bonds
  than CMC~I, fewer hydrogen bonds to water, slower hydrogen bond dynamics,
  and fewer contacts with the counterions.

%%%%%%%%%%%%%%%%%%%%%%%%%%%%%%%%%%%%%%%%%%%%%%%%%%%%%%%%%%%%%%%%%%%%%%
%%
%%
%%
%%
%%
%%%%%%%%%%%%%%%%%%%%%%%%%%%%%%%%%%%%%%%%%%%%%%%%%%%%%%%%%%%%%%%%%%%%%%

 \section{Acknowledgements}
 
 We would like to thank the John von Neumann Computer Center of the
 Forschungszentrum J\"{u}lich for Cray T90 time.  Fruitful discussions with
 Markus Deserno and Roland Faller are gratefully acknowledged.
 
 \newcommand{\fitcomment}{ The stretched exponential's {$f(x)=\exp(-(t/A)^B)$}
   integral is {$\int_0^\infty f(x) \mathrm{d}x = (A/B)\Gamma(1/B)$}.  Fits
   were performed using the nonlinear least{-}squares Marquardt{-}Levenberg
   algorithm}

 \clearpage

%\bibliographystyle{acs}
%\bibliography{cellulose,computer-md,cmc,misc,diplom,pas,local}

\end{document}